\documentclass[]{mn2e}
\usepackage{times}
\input{psfig.sty}

\newif\ifAMStwofonts

\def\Mesz{M\'esz\'aros~}
\def\Pacz{Paczy\'nski~}

\def\p{$e^\pm \;$}
\def\msun{M$_{\odot}$}
\begin{document}

\title[Merging Neutron Stars and GRBs] {High Resolution Calculations
of Merging Neutron Stars III: Gamma-Ray Bursts}
 
\author[Rosswog, Ramirez-Ruiz \& Davies]{Stephan Rosswog$^{1}$, Enrico
        Ramirez-Ruiz$^{2}$ and Melvyn B. Davies$^{1}$\\${\bf 1.}$
        Department of Physics and Astronomy, University of Leicester,
        LE1 7RH, Leicester, UK.  \\${\bf 2.}$ Institute of Astronomy,
        Madingley Road, Cambridge, CB3 0HA, UK.}

\date{}

\maketitle

\label{firstpage}

\begin{abstract}
Recent three dimensional, high-resolution simulations of neutron star
coalescences are analysed to assess whether short gamma-ray bursts (GRBs) 
could originate from such encounters. The two most popular modes of energy
extraction -- namely the annihilation of $\nu\bar{\nu}$ and
magnetohydrodynamic processes- are explored in order to investigate their
viability in launching GRBs. We find that $\nu\bar{\nu}$ annihilation
can provide the necessary stresses to drive a highly relativistic
expansion. But unless the outflow is beamed into less than one percent of
the solid angle this mechanism may fail to explain the apparent isotropized
energies implied for short GRBs at cosmological distances. We argue that
the energetic, neutrino-driven wind that accompanies the merger event 
will have enough pressure to provide adequate collimation to the 
$\nu\bar{\nu}$-annihilation-driven jet, thereby comfortably satisfying
constraints on event rate and apparent luminosity.
We also assess magnetic mechanisms to transform the available energy 
into a GRB. If the central object does not collapse immediately into a 
black hole it will be convective and it is expected to act as an effective
large scale dynamo amplifying the seed magnetic fields to a few times 
$10^{17}$ G within a small fraction of a second. The associated spindown 
time scale is 0.2 s, coinciding with the typical duration of a short GRB.
The efficiencies of the various assessed magnetic processes are high
enough to produce isotropized luminosities in excess of $10^{52}$ ergs/s
even without beaming.
\end{abstract}


\section{Introduction}

The progenitors of gamma-ray bursts (GRBs) are not well
identified. The current view of a majority of researchers is that GRBs
arise in a very small fraction ($\sim 10^{-6}$) of stars which undergo
a catastrophic energy release event towards the end of their evolution
(see \Mesz 2002 for a recent review). Observationally (Kouveliotou et
al. 1993) the short ($<$ 2 s) and long ($>$ 2 s) bursts appear to
represent two distinct subclasses, and one early proposal to explain
this was that accretion induced collapse of a white dwarf (WD) into a
neutron star (NS) might be a candidate for the long bursts, while
NS-NS mergers could provide the short ones (Katz \& Canel 1996).

Observational studies related to the possible progenitors are restricted, so 
far, to the class of long duration bursts. For these bursts, the fading X-ray
and optical afterglow emission is predominantly localized within the
optical image of the host galaxy (Bloom et al. 2002). Several of these
hosts show signs of ongoing star formation activity, necessary for the
presence of young, massive progenitor stars. This is in disagreement
with most predictions of the merger site of NS-NS binaries
which suggest that high spatial velocities and long inspiral times
would take these binaries, in more than half
of the cases, outside of the confines of the host galaxy before they
merge and produce a burst (Fryer, Woosley \& Hartmann 1999; but see
Belczynski, Bulik \& Kalogera 2002). Moreover, theoretical
estimates (e.g. Ruffert et al. 1997; Popham et al. 1998; Narayan et
al. 1992, 2001; Rosswog \& Davies 2002; Lee \& Ramirez-Ruiz 2002)
suggest that NS-NS and NS-BH mergers are only likely to produce bursts
of the short category.

According to the above reasoning, there is a possible association
between the type of GRB progenitor and the duration of the
burst. Short GRBs may involve compact binaries that should occur
predominantly in the low-density outskirts of the host galaxy while long GRBs
are more likely associated with massive stars that, at the end of
their evolution, are still within the high density cloud in which they
formed (e.g. Woosley 1993; Chevalier \& Li 1999; Ramirez-Ruiz et
al. 2001). This picture of a long GRB as a death throe of a massive
star is further corroborated by the observation of underlying supernovae
in the afterglows of several bursts, so far most convincingly demonstrated for
GRB030329 (Stanek et al. 2003).

Compact binaries are appealing candidates since they provide huge
reservoirs of gravitational binding energy and, owing to centrifugal
forces, a {\it baryon-clean} region will form naturally along the
binary rotation axis. It is in this region where a sudden energy
release is quickly and continuously transformed into a radiation
dominated fluid with a high entropy per baryon (Cavallo \& Rees 1978;
Goodman 1986; \Pacz 1986; Shemi \& Piran 1990; Piran 1999). 
In numerical simulations of such an event a high resolution of the outer 
debris layers is needed because even a tiny mass fraction in the outflow 
severely limits the attainable Lorentz factor -- for instance, a fireball of
$\sim 10^{53}$ erg would not be able to accelerate an outflow to
$\Gamma \approx 100$ if loaded with more than $\sim 5 \times 10^{-4}$
\msun. An $e^\pm-\gamma$ fireball arises from the enormous
compressional heating and dissipation associated with the accretion --
possibly involving $\nu\bar\nu$ annihilation or magnetic fields in
excess of $10^{14}$ Gauss -- which can in principle provide the
driving stresses leading to relativistic expansion. Both modes of
energy extraction from such binary encounters are discussed in this
paper.

Global simulations of compact object mergers have been performed by
various groups in order to assess their viability as possible GRB
progenitors (Davies et al. 1994, Ruffert et al. 1996, 1997, 1999,
2001, Rosswog et al. 1999, 2000, 2002a, 2002b, Lee \& Kluzniak 1999a,
1999b, Lee 2000, Lee 2001). So far it has only been possible to
address the $\nu\bar\nu$ annihilation mechanism quantitatively
(Jaroszynski 1993, Mochkovitch et al. 1993, Ruffert et al. 1997b,
Popham et al. 1999, Ruffert \& Janka 1999, Asano \& Fukuyama 2000,
2001, Salmonson \& Wilson 2001, Rosswog \& Ramirez-Ruiz 2002). 
Simulations as to how magnetised, three dimensional flows close to nuclear
density behave in strong gravitational fields are
still to be done.  The key to using simulations productively is to
isolate questions that can realistically be addressed and where 
the outcome is unknown. The analysis of such simulations then can guide
and inspire the way we think about the problem and the questions we pose.

In this paper, we report on the three-dimensional high-resolution merger
simulations, which were done using a realistic equation of state 
and emloying a detailed multi-flavour neutrino treatment. 
$\S$ 2 gives a short review of the physics and
techniques used in the current model, followed by a concise summary of
previous results. $\S$ 3 discusses the potential of $\nu\bar\nu$
annihilation as a viable source for GRB production and compares it
with several magnetic GRB mechanisms. $\S$ 4 provides a
detailed study of the distribution of merger sites with respect to
their host galaxies. The summary and a discussion of the results are
then presented in $\S$ 5.

\section{Basic model features and previous results}

We have performed a set of high-resolution simulations of the last
inspiral stages and the final coalescence of a double neutron star
system. Information regarding the numerical method, the initial
conditions and the hydrodynamic evolution can be found in Rosswog \&
Davies (2002) hereafter referred to as paper I. A detailed
account  of the involved neutrino physics and its numerical treatment
is presented in Rosswog \& Liebend\"orfer 2003 (paper II). In the
current paper we turn our attention to the viability of NS
binary coalescences as central engines of GRBs, especially of those
belonging to the short duration category.

The smoothed particle hydrodynamics method (SPH; e.g. Benz 1990 or
Monaghan 1992) is used to solve the equations of hydrodynamics for the
neutron star fluid. We have used a vastly improved artificial
viscosity scheme (Rosswog et al. 2000) for these simulations with
artificial viscosity being active only during shocks and being
efficiently suppressed in pure shear flows (see paper I for a
measurement of the effective $\alpha$-viscosity in the simulations).
We use a nuclear equation of state (EOS) for hot and dense nuclear
matter. Our EOS is based on the tables provided by Shen at al. (1998a,
1998b), the lepton and photon contributions are included, and it is
extended smoothly to the low-density regime with a gas consisting of
neutrons, alpha particles, electrons, positrons and photons. For
details concerning the EOS we refer the reader to paper I. To account
for cooling and changes in the electron fraction of the neutron star
debris we have implemented a detailed multi-flavour neutrino treatment
(paper II). We use a binary tree (Benz 1990b) to calculate the 
Newtonian self-gravity and add the back-reaction forces that
emerge from the emission of gravitational waves.

We follow the system evolution starting at an initial separation of $\sim
3R_{ns}$, where $R_{ns}$ is the radius of an isolated NS. The initial
separation is comparable to the dynamical stability limit of the
binary, so that the two neutron stars merge within a few milliseconds,
and leave behind a differentially rotating, super massive ($\sim 2.5$
\msun) NS surrounded by  dense, shock-heated debris.  The
differential rotation is expected to stabilise the central object
efficiently against gravitational collapse. Although the 
time scale until collapse sets in is quite sensitive to
the details of the merger, the remnant may remain stable for a time that
is interesting for the production of a GRB (Rosswog \& Ramirez-Ruiz 2002). 
The debris
is heated initially via shocks (for example when two spiral arms merge
supersonically, see Fig. 13 in paper I) and later via shear
motion. The innermost part of the debris torus is heated from shocks
(to $T \sim 3$ MeV) that emerge when cool, equatorially inflowing
material collides with matter being shed from the central object
leading to a butterfly-shaped temperature distribution in the
r-z-plane (paper I). The cool equatorial inflow ($T < 1$ MeV) allows
heavy nuclei to be present with a mass fraction of $\sim 10$ \%, these
nuclei typically have mass numbers $A \sim 80$ and $Z/A \sim 0.3$. The
other parts of the inner disc are essentially dissociated into
neutrons and protons. We find neutrino luminosities of $\sim 2 \times
10^ {53}$ erg/s which are dominated by electron antineutrinos. The
mean energies are $\sim 8, \sim 15, 20-25$ MeV for electron neutrinos,
electron antineutrinos and the heavy lepton neutrinos, respectively.

\section{Gamma-ray burst mechanisms}
The dominant energy from a NS-NS merger can be extracted after a black 
hole (BH) has formed, rather than during the precursor stage (Rees 1999). 
The expected outcome of such an encounter (for most presumed equations of
state) would be a rapidly spinning BH, orbited by a neutron-rich high-density 
torus. The binding energy of the orbiting debris, and the spin energy
of the BH represent the two main reservoirs. The first provides up to 42\%
of the rest mass energy of the torus, while the latter can grant up to
29\% (for a maximal spin rate) of the mass of the BH itself. 
Since the orbital velocities of an inspiralling compact binary close 
to contact reach a substantial fraction of the speed of light, the 
resulting BH is guaranteed to be rapidly spinning. This effect is more 
pronounced for softer equations of state since they yield more compact
stars. Being more massive, the BH contains an even larger reservoir of
energy than the torus. 

The central BH will have a mass of about 2.5 \msun, while the
rest of the material is either tightly bound in a torus of M$_{\rm t} 
\sim 0.15$ \msun or almost unbound, launched into highly eccentric orbits. 
A small fraction of a solar mass escapes the system. The extractable energy
could amount to several times $10^{53}({\rm M}_{\rm t}/$\msun) ergs, thus
easily satisfying the total apparent isotropic energies inferred for most
GRBs at cosmological distances. The amount of thermal and kinetic energy
of both the central object and the debris are given in Table \ref{energy_res}.
When comparing different runs the reader should keep in mind 
that they are not all calculated exactly at
the same time. They are divided into contributions from the central
object (``$>13$''; corresponding to densities above $10^{13}$
g cm$^{-3}$) and those from the orbiting debris (``$<13$''). Typically
(run C), the differentially rotating central object contains $2 \times
10^{52}$ erg and $8 \times 10^{52}$ erg in thermal and kinetic energy,
respectively. In the debris, the kinetic energy dominates by almost an
order of magnitude ($\sim 10^{52}$ versus $\sim 10^{51}$ ergs). The
corotating runs (A, B, D) yield smaller thermal energy contributions
from the central object (very smooth merger process), but
substantially more kinetic energy is found in the low density regime
(due to the larger initial angular momentum).

\begin{table*}
\caption{Energy reservoirs (in ergs) in the merged remnant.
Second column: ``cor.'': corotation, ``irr.'': no initial spin,
``$\nu+$/$\nu-$'': neutrino cooling included/not included, the number
refers to the number of SPH-particles used. ``kin'' refers to kinetic,
``therm'' to thermal energy. $>/< 13$ indicates whether densities
above (``central object'') or below $10^{13}$ gcm$^{-3}$ are
considered.}
\begin{flushleft}
\begin{tabular}{ccccccccccc} \hline \noalign{\smallskip}
run & comments & t$_{\rm end}$ [ms] & E$_{{\rm therm},>13}$ & E$_{{\rm kin},>13}$ 
& E$_{{\rm therm},<13}$ &  E$_{{\rm kin},<13}$\\ \hline \\
A   & cor., $\nu-$, $\;$ 207 918 & 10.7  &9.5 $\times 10^{51}$ & 4.6 $\times 10^{52}$ &1.8 $\times 10^{51}$ &1.9 $\times 10^{52}$ &\\
B   & cor., $\nu -$, 1 005 582 &10.8 &9.6 $\times 10^{51}$ &4.8 $\times 10^{52}$ &1.9 $\times 10^{51}$ &1.9 $\times 10^{52}$ &\\
C   & irr., $\nu +$, $\;$ 383 470 &18.3 &2.4 $\times 10^{52}$ &8.0 $\times 10^{52}$ &1.8 $\times 10^{51}$ &1.1 $\times 10^{52}$ &\\
D   & cor., $\nu +$, $\;$ 207 918 &20.2 & 1.3 $\times 10^{52}$& 4.2 $\times 10^{52}$&3.5 $\times 10^{51}$ &1.9 $\times 10^{52}$ &\\
E   & irr., $\nu +$, $\;$ 750 000&12.2 &5.3 $\times 10^{52}$ &1.4 $\times 10^{53}$ &4.7 $\times 10^{51}$ &2.4 $\times 10^{52}$ &\\
\\
F   & cor., $\nu +$, 1 005 582 &10.7 &9.6 $\times 10^{51}$ & 4.8 $\times 10^{52}$ & 1.9 $\times 10^{51}$ & 1.9 $\times 10^{52}$ &\\
\end{tabular}
\end{flushleft}
\label{energy_res}
\end{table*}

How can the available energy now be transformed into outflowing
relativistic plasma after such a coalescence event? As argued above,
we consider two options, both of which are sketched in Figure  \ref{fig}.
A fraction of the energy released as neutrinos
is expected to be converted, via collisions,
into electron-positron pairs and photons. Neutrinos can give rise to
a relativistic, pair-dominated wind if they are converted into pairs in
a region of low baryon density (Fig. \ref{fig}a). Alternatively,
strong magnetic fields anchored in the dense matter could convert the
gravitational binding energy of the system into a Poynting-dominated
outflow (Fig. \ref{fig}b). A detailed assessment of the viability of
both mechanisms follows.

\subsection{Neutrino processes}

The hot and dense merger remnant is opaque to most forms of radiation.  
Only neutrinos can extract the large thermal
energies stored in the remnant on a short time scale ($\sim 1$ s).
While they are essentially trapped inside the central object
(optical depths $\tau\sim 10^4$), they can leave the hot torus
after only a few interactions ($1 < \tau < 10$; see paper II). 
Most neutrinos are 
produced in the neutron rich ($Y_e \sim$ 0.1) debris torus around the 
central object, which exhibits temperatures well above the pair 
production threshold. The conditions in the debris favour 
positron capture on free neutrons over electron captures and therefore
electron anti-neutrinos dominate the  neutrino luminosities. For details
concerning the neutrino emission we refer to  paper II.

The neutrino emission is mildly focused around the rotation axis, 
an observer looking down on the remnant along the original binary 
rotation axis would see it around 25 times neutrino brighter than
in an edge-on observer. It is in the region above the poles of the
merged remnant, surrounded by the high-density walls of the thick disk,
where $\nu \bar{\nu}$ annihilation could accelerate a relativistic 
outflow. Neutrinos emitted from the inner shock-heated regions of the debris, 
will also deposit part of their energy in the remnant thus driving an
energetic baryonic wind (see Fig. \ref{fig}a). Let us first consider the
 $\nu \bar{\nu}$ annihilation.

\subsubsection{$\nu \bar{\nu}$ annihilation}

The annihilation of neutrinos and antineutrinos into \p as a mechanism
to power GRBs has been calculated by several groups (Jaroszynski
1993, Mochkovitch et al. 1993, Ruffert et al. 1997b, Popham et
al. 1999, Ruffert \& Janka 1999, Asano \& Fukuyama 2000, 2001,
Salmonson \& Wilson 2001). To obtain a highly relativistic outflow,
the energy from the neutrino annihilation needs to be deposited in a
baryon-poor region. The funnel above the central object of the merged
remnant is an attractive place for this deposition because not only 
is it close to the central energy source but contains --due to
centrifugal forces-- only a small number of baryons.  Unfortunately
the exact densities in this area are numerically very difficult to
resolve. There are actually no SPH-particles along the rotation axis;
the finite density estimate results from contributions of particles
located in the high-altitude parts of the debris. Due to this effect, better
resolved simulations will yield steeper density gradients towards the
z-axis. This is indeed confirmed by comparing the results from run D
and B. The density estimates along the z-axis of the merged remnant
therefore have to be considered as upper limits. The density contours 
found in the simulations are shown in the left column of Fig. \ref{rho_rz}.

We analyse the annihilation of neutrino pairs in a post-processing
approach since it is currently not feasible to perform such an
expensive calculation together with a full 3D hydro simulation.  For
the calculation of the neutrino annihilation we follow the approach
described in Ruffert et al. (1997), i.e. the following assumptions are
made: (i) relativistic effects are ignored since they are not included in the
hydrodynamics calculations either; (ii) neutrinos are
considered to be emitted isotropically into the half-space determined by
the vector $\hat{n}= -\nabla \rho /|-\nabla \rho|$; and (iii) the
particle distribution in the upper half-space of the remnant is
covered with a 3D Cartesian grid (121 x 121 x 41 grid cells) and, to
avoid additional complications due to projection effects, a grid cell of
volume $(\Delta x)^3$ is replaced by a sphere of the same volume 
(i.e.  $R= \left(\frac{3}{4\pi}\right)^{1/3} \Delta x$).  The
discretized form of the energy deposition rate per volume (erg
s$^{-1}$ cm$^{-3}$) is then given by
\begin{eqnarray}
&&Q_{\nu \bar{\nu}}(\vec{r})=\sum_{i= e,\mu,\tau} Q_{\nu_i \bar{\nu}_i}(\vec{r})\nonumber\\
&=&\sum_{i= e,\mu,\tau} A_{1,i} 
\sum_k \frac{L^k_{\nu_i}}{d_k^2} \sum_{k'} \frac{L^{k'}_{\bar{\nu}_i}}{d_{k'}^2}
[ \langle E_{\nu_i}\rangle^k +  \langle E_{\bar{\nu}_i}\rangle^{k'}] 
\mu_{kk'}^2 \nonumber\\
 &+&\sum_{i= e,\mu,\tau} A_{2,i} \sum_k \frac{L^k_{\nu_i}}{d_k^2} \sum_{k'} 
\frac{L^{k'}_{\bar{\nu}_i}}{d_{k'}^2}
 \frac{\langle E_{\nu_i}\rangle^k +  \langle E_{\bar{\nu}_i}\rangle^{k'}}
{\langle E_{\nu_i}\rangle^k  \langle E_{\bar{\nu}_i}\rangle^{k'}}
\mu_{kk'}\label{ann}
\end{eqnarray}
where the index $i$ labels the type of neutrino.  Here $L^k$ is the
neutrino luminosity of grid cell $k$, $d_k$ is the distance from the
centre of grid cell $k$ to the point $\vec{r}$, $d_k=
|\vec{r}-\vec{r}_k|$, $\langle E_{\nu_i}\rangle^k$ is the average
neutrino energy in grid cell $k$, $\mu_{kk'}= 1-\cos \theta_{kk'}$ and
$\theta_{kk'}$ is the angle at which neutrinos from cell $k$ encounter
anti-neutrinos from cell $k'$ at the point $\vec{r}$.  These relations
are sketched in Fig. \ref{vecs}.  The constants are given by
$A_{1,e}=\frac{1}{12\pi^2}\frac{\sigma_0}{c(m_ec^2)^2}
[(C_V-C_A)^2+(C_V+C_A)^2] $,
$A_{1,\mu}=A_{1,\tau}=\frac{1}{12\pi^2}\frac{\sigma_0}{c(m_ec^2)^2}
[(C_V-C_A)^2+(C_V+C_A-2)^2] $, $A_{2,e}=
\frac{1}{6\pi^2}\frac{\sigma_0}{c} [2 C_V^2-C_A^2]$,
$A_{2,\mu}=A_{2,\tau}= \frac{1}{6\pi^2}\frac{\sigma_0}{c} [2
(C_V-1)^2-(C_A-1)^2] $, where $C_V= 1/2+2 \sin^2 \theta_W$, $C_A=
1/2$, $\sin^2\theta_W= 0.23$ and $\sigma_0= 1.76 \times 10^{-44}$
cm$^2$.\\

The annihilation calculation is computationally extremely expensive
since, in principle, a loop over all {\em pairs} of grid cells is
neccesary to get the annihilation rate $Q(\vec{r},t)$ in a single
space-time point, see equation ({\ref{ann}). Since it is at present
not feasible to calculate the $\nu \bar{\nu}$ annihilation for
every time slice, we only perform it at the end of the simulation
where the neutrino luminosities have reached their final, stationary
levels (apart from maybe run C; see Figs. 2-4 in paper II). We assume
mirror symmetry with respect to the orbital plane, i.e. we only
calculate $Q(\vec{r})$ in the upper half-plane, which is a excellent
approximation for the systems considered here.

In order to further reduce the computational load,
we determine in a preprocessing step the number of grid cells that
have non-negligible contributions to the annihilation
process. Choosing for each neutrino species only those grid cells whose
luminosity is above $10^{-2}$ times the maximum luminosity (of this
specific neutrino type), we can reduce the computing time by a factor
of $\sim 10^2$ with respect to using all grid cells. The total
annihilation rates derived from both methods agree to within 1 \%. All
these post-processing calculations are performed with a fully
parallelised code.

Figure \ref{rho_rz} (right column) shows the contours of the
(azimuthally averaged) annihilation energy per time and volume, $Q$.
It is in the region along the z-axis where prohibitive baryon
loading could be avoided, transforming the $\nu \bar{\nu}$ energy
flux into a radiation-dominated fluid with a high entropy per
baryon. The thick disc geometry of the remnant with its steep density
gradients in the radial direction does not allow for lateral
expansion.  The only escape route is along the initial binary rotation
axis. The disc geometry is therefore responsible for channelling the
relativistic outflow into a pair of anti-parallel jets. This mechanism
is similar to that envisaged by MacFadyen \& Woosley (1999) for the
collapsar scenario, but offers the advantage that the jets in our case
do not have to pierce through a surrounding stellar mantle.

Figures \ref{jetC}-\ref{jetE} show the ratio of \p energy deposition
to baryon rest mass energy, $\eta=Q_{\nu\bar{\nu}}\tau_{\rm
inj}/(\rho c^2)$, in the region above the poles of the merged
remnant. $\eta$ here is calculated by using azimuthally averaged
values of both $Q_{\nu\bar{\nu}}$ and $\rho$ (in contrast to
those shown in Rosswog \& Ramirez-Ruiz 2002 which are in the xz-plane),
and can be understood as an indication of the terminal bulk Lorentz
factor $\Gamma$.  The above estimates assume, for simplicity, that the
energy deposition rate by $\nu\bar{\nu}$ annihilation into \p
pairs at the displayed time is both representative of the subsequent phases
and steady for at least one second (this is clearly an upper
limit to the total injected energy over the assumed period $\tau_{\rm
inj} \sim 1$ s). Due to the centrifugal evacuation of the funnels
the largest Lorentz factors 
are found along the binary rotation axis and therefore neutrino
annihilation will result in a pair of relativistic jets.
For our most realistic case, run C, we find peak asymptotic
Lorentz-factors of $\Gamma_{as} \approx \eta \approx 15$. 
This value is certainly resolution biased, with higher resolution 
resulting in larger Lorentz-factors. 
For example, for the corotating run D,
we find peak Lorentz-factors of $\sim 10^2$, while for the additional,
corotating run F (see Table 1), which was not included in 
papers I and II, we find a peak Lorentz-factor of 
$\Gamma_{as} \sim 10^5$ see Figure \ref{jetD}.
Corotation is certainly a somewhat artificial initial condition
(see Bildsten and Cutler 1992), but in the light of this result
we do not consider the observational constraints on the Lorentz-factors 
of GRB-outflows of several hundred (Lithwick and Sari 2001) to be a 
serious problem.
At large angles away from the rotation axis
an increasing degree of entrainment leads to a drastic decrease in
$\eta$ (see Figs. \ref{jetC}-\ref{jetE}). A broad velocity profile is
likely to develope so that different parts move with different
Lorentz factors.  This is expected to lead to a mixing instability,
which may provide the erratic changes of the Lorentz-factor that will
cause substantial irregularity or intermittency in the overall outflow
that would manifest itself in internal shocks (e.g. Rees \&
M\'esz\'aros 1994; Ramirez-Ruiz \& Lloyd-Ronning 2002)

Typical energies of $3 \times
10^{48}$ erg (run C), $2 \times
10^{47}$ erg (run D) and $2 \times
10^{49}$ erg (run E) are found in the jet
with $\eta >1$ (see Fig. 2 in Rosswog and Ramirez-Ruiz 2002). For
$\nu\bar{\nu}$ annihilation from neutron star mergers to be
the driving mechanism behind short GRBs, substantial beaming,
$\Omega_{\nu\bar{\nu}} < 10^{-2}$, is needed to satisfy the
requirements on the apparent isotropized energies of $\sim 10^{51}$
erg implied for {\it short-hard} bursts at $z=1$ (Panaitescu, Kumar \&
Narayan 2001; Lazzati et al. 2001). Possible beaming mechanisms are 
discussed below.

\subsubsection{Neutrino-driven winds}

A fraction of the neutrino energy emitted from the inner regions of
the debris will be deposited in the other parts of the remnant
(Fig. \ref{fig}a), heating and driving material from its surface
(Woosley 1993; Ruffert et al. 1997). For example, a nucleon located at
$\sim 100$ km from the central object gains the equivalent of its
gravitational binding energy by capturing 3 neutrinos of $\sim 15$
MeV.  Using typical numbers from our simulations, we find that the
neutrinos will drive a mass outflow at a rate (Qian \& Woosley 1996):
\begin{equation}
\dot{M}\approx 2 \times 10^{-2} {\rm M}_{\odot}/{\rm s} \;\;
L_{{\bar{\nu}_e},53}^{5/3} 
\left(\frac{\epsilon_{\bar{\nu}_e}}{15 {\rm MeV}}\right)^{\frac{10}{3}}  
\left(\frac{R_{\rm em}}{80 {\rm km}}\right)^{\frac{5}{3}}
\left(\frac{2.5 {\rm M}_{\odot}}{M_{\rm em}} \right)^2 \nonumber,
\end{equation}
where $L_{\bar{\nu}_e}$ is the luminosity in
$\bar{\nu}_e$-neutrinos, $\epsilon_{\bar{\nu}_e}=
\frac{\langle E^2_{\bar{\nu}_e}\rangle}{\langle
E_{\bar{\nu}_e} \rangle}$, $E_{\nu}$ denotes the neutrino energy,
$R_{\rm em}$ and $M_{\rm em}$ are the radius of and the mass enclosed
in the neutrino emitting region.  This estimate is consistent with
those found by Woosley (1993) and Duncan, Shapiro \& Wassermann
(1986). This wind accelerates the blown-off material to its asymptotic
velocities of a few $10^9$ cm s$^{-1}$.

\subsubsection{Hydrodynamic collimation}

As shown above, the question of whether or not $\nu\bar{\nu}$ 
annihilation is a viable candidate for driving GRBs depends on 
the possibility of collimating the plasma into a narrow beam. 
We will briefly summarize a beaming mechanism related to the 
neutrino-driven wind (beaming by {\it hoop stresses} produced by the 
toroidal component of a magnetic field configuration will be addressed later).
The wind mechanism is based on the idea of Levinson \& Eichler (2000)
that the pressure and inertia of a baryonic wind can lead to the collimation
of a baryon-poor jet, the detailed assessment of the wind-jet
interaction for the neutron star merger case is worked out in Rosswog 
\& Ramirez-Ruiz (2003).
The main result is that bursts produced by such a mechanism are not isotropic 
but instead beamed within a solid angle, with opening half-angles of
\begin{equation}
\theta \approx {\pi \over \beta_w} {L_j \over L_w} \approx
0.3\;\beta_{w,-1}^{-1}\;L_{w,50}^{-1}\;L_{j,48.5},
\end{equation}
and hence the apparent luminosity is
\begin{equation}
L_\Omega \approx {2 \over \pi^2} {\beta_w^2 L_w^2 \over L_j} = 2
\times 10^{51} \beta_{w,-1}^{2} L_{w,51}^{2}\;L_{j,48}^{-1} \;{\rm erg\;
s^{-1}},
\end{equation}
where $\beta_w$ is the wind terminal velocity in units of the speed of light, 
$L_w$ and $L_j$ are the 
total power in the wind and in the jet, and we adopt the convention 
$Q = 10^x\,Q_x$, using cgs
units. The proportionality of the opening angle on the jet luminosity
yields the paradoxical result that, if all other quantities are the same,
a low-luminosity jet will appear brighter than a high-luminosity one.
We find for our generic run C  $\theta \approx 0.1$ and
$L_\Omega \approx 10^{51}$ erg s$^{-1}$. With this beaming fraction
the BATSE detection rate of short GRBs yields a required event rate
that comfortably falls into the range of estimated neutron star
merger rates (Rosswog \& Ramirez-Ruiz 2003). 
We find that even a relatively narrow distribution of neutron star
masses yields a broad distribution of opening angles and apparent 
luminosities. These distributions can be found in Rosswog \& 
Ramirez-Ruiz (2003). This model leads to a simple physical interpretation 
of the isotropic luminosities implied for short GRBs at cosmological
distances.

If we use the found energy per solid angle 
and compare it to that of
long bursts (Panaitescu \& Kumar 2001) we find that the afterglow of
short bursts produced in this way should be typically a factor of 50-100
dimmer than those of long bursts (the afterglow energy is roughly 
proportional to the energy in the outflow). If short GRBs arise from
NS-NS or NS-BH binaries merging in the tenuous outskirts of their host
galaxies then our chances for detecting radio and optical afterglows
of short GRBs are further diminished since the afterglow brightness in
these bands is proportional to the square root of the external
density. Assuming that the relativistic shocks in long
and short GRBs have similar parameters (i.e. p=2, magnetic field 
is 5 \% of equipartition) and using a total isotropic energy
of $10^{51}$ erg and a medium density of $n=10^{-3}$ cm$^{-3}$ we can 
calculate the 2-10 keV flux (for details see Ramirez-Ruiz, Celotti \& Rees
2002). This is shown in Fig. \ref{xmm} as solid line. 
 The best chance of detecting afterglows of short GRBs is
with early X-ray observations with {\it XMM} and {\it Chandra} within
a few days after the GRB. These conclusions are illustrated in Figure
\ref{xmm}, which shows the X-ray emission expected for a short GRB
afterglow with lower values of $E\approx 10^{51}/4\pi$ erg s$^{-1}$
and $n\approx 10^{-3}$ cm$^{-3}$ than those found for long duration
ones ({\it solid line}). Also shown are the {\it Chandra} upper limits
for GRB 020531 (Butler et al. 2002)
and GRB 021201 
(Hurley et al. 2002), for which the afterglow vanished too soon to
permit detection.

The question is why these bursts have been detected in $\gamma$-rays but
not in X-rays or optical, which would have been expected if the X-ray
to $\gamma$-ray ratio had been comparable to those in long duration
GRBs. One simple reason may be that the afterglows of short GRBs
are, on average, dimmer than those of long bursts. This speculation is
confirmed by the transient and fading hard X-ray emission obtained by
summing up the light curves of all the brightest short bursts seen by
{\it BATSE} (Lazzati et al. 2001).  Extrapolation of such a signal to
later times shows that the typical short GRB afterglow is close to the
current detection limits ({\it grey region} in Fig. \ref{xmm}). We
note that there are short GRBs with fluences significantly larger than
average. These bursts could be as energetic as some long GRBs for
which afterglows have been seen (and may contribute significantly to
the average signal shown in Fig. \ref{xmm}). The afterglows of such
short GRBs could be detected beyond a day with {\it XMM}.

\subsection{Magnetic processes}

In this paragraph, we discuss some general ways by which gravity, 
angular momentum and electromagnetic field could conspire to power 
GRBs (see e.g. Narayan et
al. 1992, Thompson and Duncan 1993, Usov 1992, 1994, M\'esz\'aros \&
Rees 1997, Katz 1997; for a more complete bibliography see the reviews
of Piran 1999 and M\'esz\'aros 2002).
There is considerable literature on the launching of
electromagnetically driven jets from the surface of an accretion disc,
under the assumption that the disc is threaded by a suitably
configured poloidal magnetic field (e.g. Blandford \& Payne 1982) but
with little discussion of how the field configuration was set up in
the first place.

The merger remnant with its large magnetic seed fields and its turbulent
fluid motion will give rise to a plethora of electromagnetic activity.
We expect coronal arches, as well as larger scale magnetic structures
to be quite common and to be regenerated on an orbital
timescale. Field lines will be stretched across the disc surface and
will quickly be forced to reconnect. Differential rotation will cause
loops to twist and probably to undergo some topological
rearrangement (Fig. \ref{fig}b). This provides an alternative mechanism
for heating the disc corona and perhaps for driving an outflow through
thermal heating (similar to the neutrino heating mechanism described
above). Starting with the typical fields of a neutron star the 
turbulent flow and the differential rotation
within the remnant will quickly increase the magnetic field to
enormous strengths. The disc may be threaded by a vertical
field and differential rotation will make the field approximately
axisymmetric. Not only will this create a very effective coupling
between the disc and some possible outflow but it can also facilitate the
outflow by launching it either centrifugally or through magnetic
pressure associated with coiled up toroidal field. There is also a
possible magnetic connection between the disc and the hole (e.g.
Blandford \& Znajek 1997; hereafter B-Z) which can be used to extract
the energy stored in the rotation of space-time. But even before the BH 
forms, a NS-NS merging system might lead to winding
up of the fields. This field amplification will continue until the
magnetic pressure becomes comparable to the gas pressure at which
point the magnetic field becomes buoyant, floats up and forms a
magnetically structured corona above the merger remnant.\\
We will now assess several mechanisms to produce a GRB in more detail.

\subsubsection{Energy extraction from the central, supermassive neutron star}  

In the following we will assume that the central object
remains stable against gravitational collapse for the time scale that 
is of interest for the discussed mechanism. We regard differential rotation
to be responsible for this stabilization and expect a large range of possible 
lifetimes (see discussion in Rosswog \& Davies 2002 and Rosswog \& 
Ramirez-Ruiz 2002).\\

\centerline{\underline{The DROCO mechanism}}

\noindent Kluzniak and Ruderman (1998) have suggested Differentially Rotating 
Collapsed Objects, so-called DROCOs, as central engines of GRBs.
The DROCO will  wind up initial seed magnetic fields by differential rotation
until the magnetic pressure, $p_{\rm mag}= B^2/8 \pi$ becomes comparable to
the matter pressure, i.e. $B \approx B^{\rm eq}$. At this point the 
magnetic toroids created by winding up magnetic fields around the rotation 
axis (the Taylor-Proudman steady state) will become buoyant, float up and 
finally break through the surface where they will create an ultrarelativistic blast.
 This mechanism -winding up of the field lines, buoyancy and 
breaking through the surface- will continue until either the kinetic energy
of the DROCO is used up or the central object undergoes gravitational 
collapse to a black hole. In this way a series of erratic, explosive 
releases of energy (``sub-bursts'') is created.\\
The hot, supermassive ``neutron star'' in the center of merger remnant
is rapidly  differentially rotating with periods ranging from 0.3 ms 
to 2 ms (see Figure 10 in Rosswog \& Davies 2002) and therefore one
such realization of a DROCO. The local saturation field amplitude is 
determined by a balance between nonlinear growth, dissipative processes 
like reconnection and buoyant
escape and can clearly only be estimated through careful, three
dimensional, numerical simulations which are just now becoming
possible. Ideally, this process should be treated with general
relativistic MHD, but lacking the corresponding tools we will use here
a simple toy model together with numbers inferred from our purely
hydrodynamic simulations to get a first handle on the possible
magnetic properties of the remnant.  We will only
consider the simplest case of field amplification, the winding up of
field lines by differential rotation (a detailed discussion of
differential rotation and stellar magnetic fields can be found, for
example, in Spruit 1999).\\
In the left panel of Fig. \ref{B_equipartition} we plot the equipartition field
strengths of the matter distribution of our generic case, run C (at
14.11 ms),
\begin{equation}
B^{\rm{eq}}= \sqrt{8 \pi \rho c_s^2}, \label{B_eq}
\end{equation}
where we have used azimuthally averaged values for the densities and
the sound velocity, $c_s$, which we derived from our nuclear EOS. The
equipartition field distribution for the other cases is very similar
and is therefore not shown. We find values between $10^{16}$ and
$10^{18}$ G in the central object ($r_{\rm cyl} < 30$ km), while the
inner parts of the dense torus ($r_{\rm cyl} < 130$ km) exhibit values
between a few times $10^{14}$ and $10^{16}$ G. The rapidly expanding
low-density regimes ($r_{\rm cyl} > 130$ km) still yield values well
above $10^{13}$ G.
Most of this material moves on eccentric orbits around the centre of
mass and will later rain down on the inner parts of the disc thereby
refuelling via shock heating the thermal energy reservoir of the
torus.\\ 
Interactions between the fluid motions $\vec{v}$ and the magnetic fields
$\vec{B}$ are governed by the induction equation.
If magnetic diffusion is unimportant for the time scales
considered here, it reads
\begin{equation}
\partial_t \vec{B}= \nabla \times (\vec{v} \times \vec{B}).
\end{equation}
If we further assume that the motion of the material can be approximated
by an axisymmetric, purely azimuthal motion, the poloidal field component,
$B_p$, does not change in time, but the azimuthal component, $B_{\phi}$ 
is increased by ``winding-up'' according to 
\begin{equation}
B_\phi= 2 \pi N(t) B_p,
\end{equation}
where $N(t)$ is the number of revolutions performed as a function of time.
If we assume for simplicity the rotation rate to be stationary on the time 
scale of interest (in reality magnetic braking will occur), we can write 
the time it takes $B_\phi$ to reach equipartition as 
\begin{equation}
\tau^{\rm{eq}}= \frac{B^{\rm{eq}}}{B_p} \omega(r)^{-1}.\label{tau_eq}
\end{equation}

We evaluate equation (\ref{tau_eq}) for our generic case, run C, at a
time where the inner parts of the remnant (central object +
high-density part of the disc) have reached a state of stationary
rotation (t= 14.11 ms).  This is shown in the right panel of Figure
\ref{B_equipartition} as a function of radius where we have used a typical 
neutron star magnetic field as initial seed, $B_p=10^{12}$ G.
The curve reflects the interplay between the radial dependencies of
$\omega$ and $B^{\rm{eq}}$. Under the above assumptions the central 
object will reach equipartition within a few 10 s, the high-density
torus within $\sim$ 4 s.\\
Starting from a seed field of 10$^{12}$ G our
simple model for the field evolution predicts that it takes around 160 
revolutions or 0.3 seconds for the central object  to reach a typical 
magnetar field strength of $10^{15}$ G. 
Equipartition, a few times $10^{17}$ G, will be reached after a few 10 s, 
the exact time depending on the position within the central object. 
At this point strongly magnetised flux 
tubes will float up at roughly the Alfven speed, $v_A= B/ \sqrt{4 \pi \rho}$
on a time scale $\tau_b \sim R_{co}/v_A \approx 1.5 \times 10^{-4}$ s.
This short rise time is consistent with observed GRBs that reach
their peak luminosity within milliseconds.
Kluzniak and Ruderman estimate the energy of a typical sub-burst to be
$\sim 10^{51}$ ergs. Using this number and the kinetic energies in the 
central objects, a few times $10^{52}$ ergs (Table \ref{energy_res}), the 
central object could launch several tens of such sub-bursts.\\

\centerline{\underline{Convective motion, dynamos and superpulsars}}

\noindent In the previous section only the relatively slow wrapping of
field lines was considered. An effective $\alpha-\omega$ dynamo
(Thompson and Duncan 1993) or a magnetorotational instability 
(Balbus and Hawley 1991, 1998), however, might amplify an initial seed
field exponentially on a much shorter time scale.\\
Figure \ref{convection} shows the velocity field in the interior 
of the central object of run C. It shows ``convective cells'' of a 
scale $l_c \sim 1$ km and convective velocities of 
$v_{\rm con} \sim  10^8$ cm/s
created by fluid instabilities in the course of the merger.
Moreover, the neutrino optical depth drops steeply from $\sim 10^4$ in the 
interior towards the edge of the central object (see Figure
11 in Rosswog and Liebend\"orfer, 2003). These outer layers lose neutrinos,
and therefore lepton number and entropy, at a much higher rate than the 
interior, thereby creating negative entropy and lepton number gradients 
(for the neutrino luminosity per volume see Figure 6, second panel, Rosswog 
and Liebend\"orfer, 2003), which, in turn, will drive vigorous convection
(e.g. Epstein 1979, Burrows and Lattimer 1988). 
Apart from being roughly twice as  massive and more neutron rich 
($Y_e \sim 0.1$ rather than $\sim 0.3$; since the individual neutron stars
have been deleptonized in the cooling phase after their formation) this
state resembles closely the convection phase after the birth of a 
protoneutron star in a type II supernova. If we assume neutrinos
to be the dominant source of viscosity, $\nu_{\nu}= 10^{9}
\cdot f(y_p) \cdot T_{15} \cdot \rho_{14}^{-4/3}$ cm$^2$/s (Thompson and Duncan 1993)
with $f(y_p)= [y_p^{1/3} + (1-y_p)^{-1/3}]^{-1}$, $T_{15}= T/15$ MeV, 
$\rho_{14}=\rho/10^{14}$ gcm$^{-3}$ and $y_p$ the proton fraction,
the Reynolds number inside the central object is 
$Re= v_{\rm con} l_c / \nu_{\nu} \sim$ few $10^4$.
The related viscous damping time scale for the convective motion is 
$\tau_c \sim l_c^2/\nu_{\nu} \sim 60$ s, corresponding to several $10^5$ 
neutron star dynamical time scales.\\
Whether such a neutron star can support a dynamo
is determined by the Rossby number (Duncan and Thompson 1992), 
$Ro \equiv T_{\rm rot}/\tau_{\rm con}$,
where $T_{\rm rot}$ and $\tau_{\rm con}$ are the rotation period and 
the convective overturn time, respectively. 
An effective large scale helical dynamo can only be supported
for $Ro < 1$. With our typical convective velocities and cell sizes
the convective overturn 
times are $\tau_{\rm con} \approx 3$ ms. The rotational periods
in the central object are $0.3 < T_{\rm rot} < 2$ ms (see Figure 10 in Rosswog and
Davies 2002), yielding
Rossby numbers $0.1 < Ro < 0.7$. We therefore expect a very strong
dynamo to be operating. Using the kinetic energy of the central object of our
generic run C, $E_{\rm kin}= 8 \cdot 10^{52}$ erg (see Table \ref{energy_res}),
we find that enough energy is available for an average field strength
$\langle B \rangle_{\rm co}= \sqrt{3 \cdot E_{\rm kin}/ R_{\rm co}^3}
\approx 3\cdot 10^{17}$ G, where $ R_{\rm co}$ is the radius of the central 
object. This is in good agreement with the equipartion field strengths shown 
in Figure \ref{B_equipartition}.\\
Such a convective dynamo amplifies seed magnetic fields exponentially
with an e-folding time, $\tau$ close to the convective time scale
(Nordlund et al. 1992). Using the numbers from our simulation equipartition 
will be reached in the central object after only 
$\tau^{eq}= \tau \ln(B^{eq}/B_0) \approx 38 $ ms.\\
Such an object could act like the ``superpulsar''
envisaged by Usov (1992). The kinetic energy from the deceleration of the 
central object is released in the form of an (initially) optically thick
electron positron wind with frozen-in magnetic field. Once the 
magnetohydrodynamic approximation breakes down far from the pulsar, 
electromagnetic waves are generated that accelerate outflowing particles to
ultrarelativistic Lorentz factors (10$^6$ or more; Usov 1994).\\
Using typical numbers from the simulation and $B= \langle B \rangle_{\rm co}
=3 \cdot 10^{17}$ G as estimated above the magnetic dipole luminosity is
\begin{equation} 
L_{\rm md}= 5.1 \cdot 10^{53} \; {\rm erg/s} 
\left( \frac{B}{3 \cdot 10^{17} \; {\rm G}} \right)^2 
\left( \frac{R}{15 \; {\rm km}} \right)^6
\left( \frac{\omega}{3000 \; {\rm 1/s}} \right)^4
\end{equation} 
and the spin down time scale will be $\tau_{\rm sd}=  E_{\rm kin}/L_{\rm md} 
\approx 0.2$ s, which maybe not just by chance coincides with the typical 
duration of a short GRB. 

\subsubsection{Energy extraction from a central black hole: the Blandford-Znajek mechanism}

Although the time scale is uncertain, the central part of the merger
remnant will sooner or later collapse to a rotating BH (unless nuclear
physics allows the support of cold objects of 2.8 \msun).  If
magnetic fields of comparable strength to those discussed above thread
the BH, its rotational energy of
\begin{equation}
E_{\rm rot}= M_{\rm bh} c^2 \left(1-\sqrt{1/2
\left(1+\sqrt{1-a^2}\right)}\right) \approx 1.5 \times 10^{53} \; {\rm
erg},
\end{equation}
can be extracted via the B-Z process (Blandford and Znajek 1977). 
Here $a= Jc/GM^2$ is the BH rotation parameter.
The above estimate has been calculated using typical numbers from our
simulations ($M_{\rm bh} = 2.5$ \msun and $a= 0.5$).  The relevant
field component for the energy extraction via the BZ-process is the
poloidal field. The ratio between the poloidal and azimuthal field is 
given by $B_p/B_{\phi} \approx (H/R)$ (Meier 2001). For the thick disks under
consideration $H \sim R$ (see Fig. \ref{rho_rz}, left panels) and therefore
we use $B= B_p \approx B_{\phi}$.  
If the BH is threaded by the equipartition magnetic field of the disc
$\approx 3 \times 10^{15}$ G then a jet with luminosity
$L_{\rm jet} = B^2H^2R^2 \Omega^2/32c \approx 10^{53}$ erg s$^{-1}$ 
(Meier 2001) will result. Again, the typical time scale will be of the order
of a typical short GRB duration: 
$\tau_{\rm BH}= E_{\rm rot} / L_{\rm jet} \approx 1.5$ s.

\subsubsection{Energy extraction from the disk}

Very similar processes are expected to take place in the disc
(Narayan et al. 1992). After around 4 s the field strength has reached
equipartition, $\sim 10^{15}$ G, and the field will float up to the disk
surface in 
$\tau_b \sim H /v_A \approx 2 \times 10^{-2}$ s where it will form a
magnetically structured disk corona.  Livio et al. (1999) estimate a
maximum possible disc luminosity (Poynting flux or magnetically driven
wind) given by
\begin{equation}
L_{\rm disk}^{\rm max} \sim \frac{B^2}{4} R^3 \omega \approx 5 \times 10^{52}
{\rm erg\;s^{-1}},
\end{equation}
where we have used the numerical values found at the outer edge of
the thick debris disk (see Fig. \ref{B_equipartition}) and assumed that both the
poloidal and azimuthal fields are of comparable strength. Even
allowing for only $ 10$ \% of this luminosity to be converted into an
MHD jet, a system powered by the disc binding energy would --without
beaming-- satisfy the apparent isotropized energy of $10^{51}$ ergs
implied for short-hard bursts at $z\approx 1$ (Panaitescu et al. 2001;
Lazzati et al. 2001). The disk would allow an energy extraction
on a possibly longer time scale than the central object, 
$\tau_{\rm disk}= E_{\rm disk} / L_{\rm disk}^{\rm max}
> 0.42 \cdot {\rm M}_{\rm disk} c^2 / L_{\rm disk}^{\rm max}
\sim 3$ s.\\

\section{Distribution of neutron star mergers with respect to their 
host galaxies}
\medskip

The distribution of neutron--star mergers with respect to their host
galaxies depends on the distribution of the times it takes from the
formation of the system to coalescence and, of course, on the galactic
potential.  All formation channels seem likely to produce binaries
receiving kicks of order 100 -- 200 km/s.  If the median merger time
is approximately $10^6$ yr, then a large fraction of mergers will
occur close to star--formation sites (i.e. within the host
galaxy). If, however, the median merger time is closer to $10^8$ yr,
then a large fraction of mergers may occur somewhat outside their host
galaxies (which are typically dwarf galaxies). Searches for host
galaxies of gamma--ray bursts may thus help limit the possible
progenitors (see e.g. Bloom et al. 2002) {\it providing} we have a
reasonable understanding of the distribution of merger times and
accurate galactic models.

\medskip

The merger time due to gravitational radiation is a sensitive function
of separation $a$ , $\tau_{\rm gr} \propto a^4$, and thus slight changes in
the initial distribution of separations  will lead to much
larger variations in the distribution of merger times. There are a
number of recent papers concerning the evolutionary pathways to
producing NS-NS binaries. Bloom et al. (1999) use the binary
evolutionary code of Pols and Marinus (1994) and produce a population
with the merger time peaking at about $10^8$ yr. Fryer et al. (1999)
consider three distinct routes to producing NS--NS binaries and find
$\tau_{\rm gr} \sim 10^8 - 10^9$ yr. More recently, Belczynski et
al. (2002a, b, c) argue that a large fraction of mergers should occur 
on much shorter time scales
($\tau_{\rm gr} \sim 10^5 - 10^6$ yr). The penultimate stage in
producing a NS--NS binary consists of one neutron star in a tight
binary with a helium star. The latter will produce the second neutron
star when it explodes as a type II supernova. However, if the helium
star expands to fill its Roche lobe, the subsequent mass transfer may
alter the separation of the two stars and thus affect the distribution
of initial separations (and therefore merger times) for the NS--NS
binaries.  Belczynski et al. argue that this mass transfer produces a
common envelope phase, where the neutron star and the core of the
helium star spiral together whilst ejecting the helium--star
envelope. For typical parameters this process will produce NS--NS
binaries having initial separations $\sim 0.2$ to $ \sim 2$
R$_\odot$. Such a population is shown in Figure \ref{mbd_fig1} for
$\alpha_{\rm ce} \lambda_{\rm ce}= 1$, where $\alpha_{\rm ce}$ is the
common envelope efficiency, $\lambda_{\rm ce}$ is a function of the
internal structure of the star and defined as in equation (9) of
Davies et al. (2002).\\ We argue, however, that there is a problem
with this scenario as a common envelope evolution would require very
large common envelope efficiencies to reproduce observed systems. In
Figure \ref{mbd_fig1} we have also plotted the WD-NS binary
J1141-6545.  This binary was produced in a manner similar to NS--NS
binaries (Davies, Ritter \& King 2002). According to the scenario
suggested in Davies et al. (2002), mass transfer occurs when the
helium star fills its Roche lobe but rather than forming a common
envelope, the material is ejected by the white dwarf, in a manner
similar to the evolution assumed for the X--ray binary Cygnus X--2
(Kolb et al. 2000).  If instead we assume mass transfer does produce a
common envelope phase, J1141-6545 can only be produced if we assume an
extremely high common envelope efficiency, $\alpha_{\rm ce}
\lambda_{\rm ce}= 10$, as shown in Fig. \ref{mbd_fig2}. But with this
high value of $\alpha_{\rm ce}$ the scenario of Belczynski et al. also
produces initial separations close to the results found by other
workers, leading to median merger times of $\sim 10^8$ yr. In fact,
Ivanova et al. (2002) suggest an alternative treatment for the mass
transfer, and also produce merger times of $\sim 10^8$ yr.

The distribution of merger times is shown in Figure \ref{mbd_fig3}
(solid lines from left to right): the merger timescales envisioned in
Belczynski et al. (2002a, b, c), those derived through
J1141-6545--like evolution, those derived neglecting the radius of the
helium star, and those derived for 2303--like evolution, where the helium 
star fails to fill its Roche lobe and instead mass-loss occurs via a wind.
 The dotted
line is for the evolution as described in Belczynski et al. (2002a, b)
 but with $\alpha_{\rm ce} \lambda_{\rm ce} = 10$ which is required
under this scenario to produce the observed system J1141-6545. Figure
\ref{mbd_fig3} shows the importance of considering the helium star
radius. Those systems filling their Roche lobes produce very tight
binaries which have short merger times. But the observed properties of
J1141-6545 impose a very important constraint on any evolutionary
scenario. With this constraint, all models produce median merger times
of about $10^8$ yrs.

We want to point out here that the agreement of theoretical
distributions of merger sites with observations is a very sensitive
function of the assumed galactic potential. Assuming a
J1141-6545--like evolution, we performed a monte carlo simulation of
the population of neutron star binaries, integrating their motion
within a galactic potential and observing their projected distances
from the galactic centre when they merge. This is shown in Fig.
\ref{mbd_fig4} for three galactic models: models d, b, and e. 
Models d and b have galactic disc lengthscales of 1 kpc, 
whereas model e has a galactic disc lengthscale of 3 kpc.
The halo scalelengths are 3kpc for all three models. Model b has
a total mass of $0.278 \times 10^{11}$ M$_\odot$ whereas models
d and e have a total mass of $0.625\times 10^{11}$ M$_\odot$
(see Bloom et al. 1999 for further details).  We also
plot the {\em observed} distribution from Bloom et al.  (2002). The
probability that the observed distribution matches the theoretical
curves is a sensitive function of galactic potential (model d: 9 \%,
model b: 4 \%, model e: 0.1 \%), i.e. just changing the mass of a
galaxy by a factor of 2 has a drastic effect on the resulting
probabilities. It should be recalled that the observed distribution 
is for a sample of long GRBs. If the merger of two compact objects 
produce short bursts, the lack of agreement between the theoretical 
curves of mergers and observations of long bursts is not a surprise.

\section{Summary and discussion}

We have addressed the viability of binary neutron star coalescences as
central engines of GRBs.  Two possible forms of outflow have been 
considered: (i) a pair dominated fireball generated by $\nu\bar{\nu}$ 
annihilation, and (ii) electromagnetic Poynting flux. 

We find that the plasma produced by the annihilation of $\nu\bar{\nu}$ pairs 
will be channeled by the high-density walls of the thick disk of the 
remnant into relativistic, bipolar jets along the original binary 
rotation axis. In extreme cases we find peak asymptotic Lorentz-factors
of up to $\sim 10^5$.
The energy contained in these jets, however, falls short of 
reproducing the isotropized energy requirements, $E_{\rm iso}\sim 10^{51}$ 
ergs for short bursts at z$\approx 1$ that are deduced from observations. 
Since we 
find typical energies of $\sim 10^{48}$ ergs only, $\nu\bar{\nu}$ the emission
has to be beamed into a small fraction of the sky in order for $\nu\bar{\nu}$ 
annihilation to be a viable GRB mechanism.
By applying models developed in the context of protoneutron stars
produced in core-collapse supernovae we find that the huge neutrino
emission of the merger remnant ($\sim 2 \times 10^{53}$ erg/s) will
drive a strong baryonic wind with mechanical luminosities of $\sim
10^{50}$ erg s$^{-1}$. This outflow will have enough pressure to
provide adequate collimation, satisfying both the isotropized energy
requirements and the estimated rate of binary mergers (Rosswog \& Ramirez-Ruiz 
2003). A second feature of the jet-wind interaction is that there will be 
some entrainment of the electron-ion plasma and this should show up in the
polarization observations, which can, in principle, distinguish a pair 
plasma from a protonic plasma.  Due to  entrainment, the jet will develop
a velocity profile with velocities decreasing away from the jet axis. 
This implies that an observer is likely to infer a
value of the Lorentz factor that depends upon the inclination of the
line of sight to the jet axis.  Seen from one viewing angle a burst
would appear as a X-ray rich burst (Heise et al. 2001) while from a different
angle it would be interpreted as a standard GRB (Salmonson 2000; 
Ioka \& Nakamura 2001; Kobayashi, Ryde \& MacFadyen 2002; 
Ramirez-Ruiz \& Lloyd-Ronning 2002; M\'esz\'aros et al. 2002).
In some cases mixing instabilities at the jet walls could spoil
the jet with baryons and result in a non-relativistic, ``dirty fireball'' 
rather than a proper GRB fireball. 
We predict the afterglow of such a burst to be dimmer than those of 
long bursts by a factor of $\sim 50-100$. The best chances to detect such 
an afterglow are with $XMM$ and $Chandra$ within a few days after the GRB.\\

An alternative way to tap the torus energy is via magnetic fields.
The large seed magnetic fields of the individual neutron stars will collude
with the complex fluid motion in the merger remnant to produce enormous 
field strengths. This may proceed via pure differential rotation, dynamo
action and/or magnetorotational instabilties.\\
If the central object resulting form the merger does not collapse
immediately, it will be subject to vigorous convection. This convection is
driven by large entropy and lepton number gradients established via 
neutrino emission and will resemble the convection in a newborn
protoneutron star.
We find Rossby numbers well below unity and therefore expect the 
central object to act as an efficient large scale dynamo amplifying
initial seed fields exponentially. We find that enough energy is 
available to build up field strengths of several times $10^{17}$ G. The 
spindown timescale of a ``neutron star'' endowed with such a field is
$\sim 0.2$ s which coincides with the average duration of a short GRB.\\
We have investigated several magnetic mechanisms to produce a GRB, 
considering the extraction of available energy from the supermassive 
neutron star in the center of the merger remnant, from a black hole that 
is expected to form at some stage after the merger and from the debris disk.
Most of these (apart from perhaps
the DROCO mechanism) seem to be able to extract the available energies
at rates in excess of $10^{52}$ ergs/s. Consuming the available energy 
reservoirs at such rates naturally leads to durations of order one second.
These mechanisms are able to satisfy the requirements on the equivalent
isotropized energies of short GRBs even without beaming.\\
Moreover, such magnetic mechanisms are supported by the recent
observation of linear polarization in the prompt gamma-ray emission
from GRB021206 (Coburn and Boggs, 2003) which suggests that the 
GRB explosion is powered by a central engine with a strong, large-scale
magnetic field.\\
Given the plethora of different mechanisms to transform the large
gravitational binding energy into gamma-rays, the enormous observed
diversity of GRBs seems to be a natural consequence. Several of the discussed
mechanisms may be active at the same time. For example, 
the directed, relativistic
jets produced via neutrino annihilation may be accompanied by uncollimated
gamma-rays produced in energetic disk flares.\\

Finally we want to add a word of caution concerning the comparison of
theoretical merger site distributions with observed ones.  We have
reexamined the distribution of merger sites with respect to their host
galaxies since this distribution is generally regarded as an important
clue the nature of the GRB progenitors. Our Monte Carlo simulations
yield median merger times of $\sim 10^8$ years rather than the much
shorter times claimed recently and are thus in agreement with several
previous results.  Using our distribution of merger times we have
calculated the projected distances of the merger sites to the galactic
centre. We found median distances of $\sim 3 $ kpc, but stress
explicitly that the probability that theoretical distributions match
observations is very sensitive to the properties assumed for the host
galaxy.

\section*{Acknowledgements}
It is a pleasure to thank M. J. Rees and G. Wynn for stimulating discussions
and the Leicester supercomputer team S.  Poulton, C. Rudge and R. West
for their excellent support.  The computations reported here were
performed using both the UK Astrophysical Fluids Facility (UKAFF) and
the University of Leicester Mathematical Modelling Centre's
supercomputer. S.R. gratefully acknowledges a PPARC Advanced Fellowship, 
E.R. was supported by CONACyT, SEP and the ORS foundation. M.B.D.
gratefully acknowledges support of a URF from the Royal Society.
Theoretical astrophysics at Leicester is supported by a PPARC rolling grant.


\clearpage



\begin{figure}
\hspace*{2cm}\centerline{\psfig{file=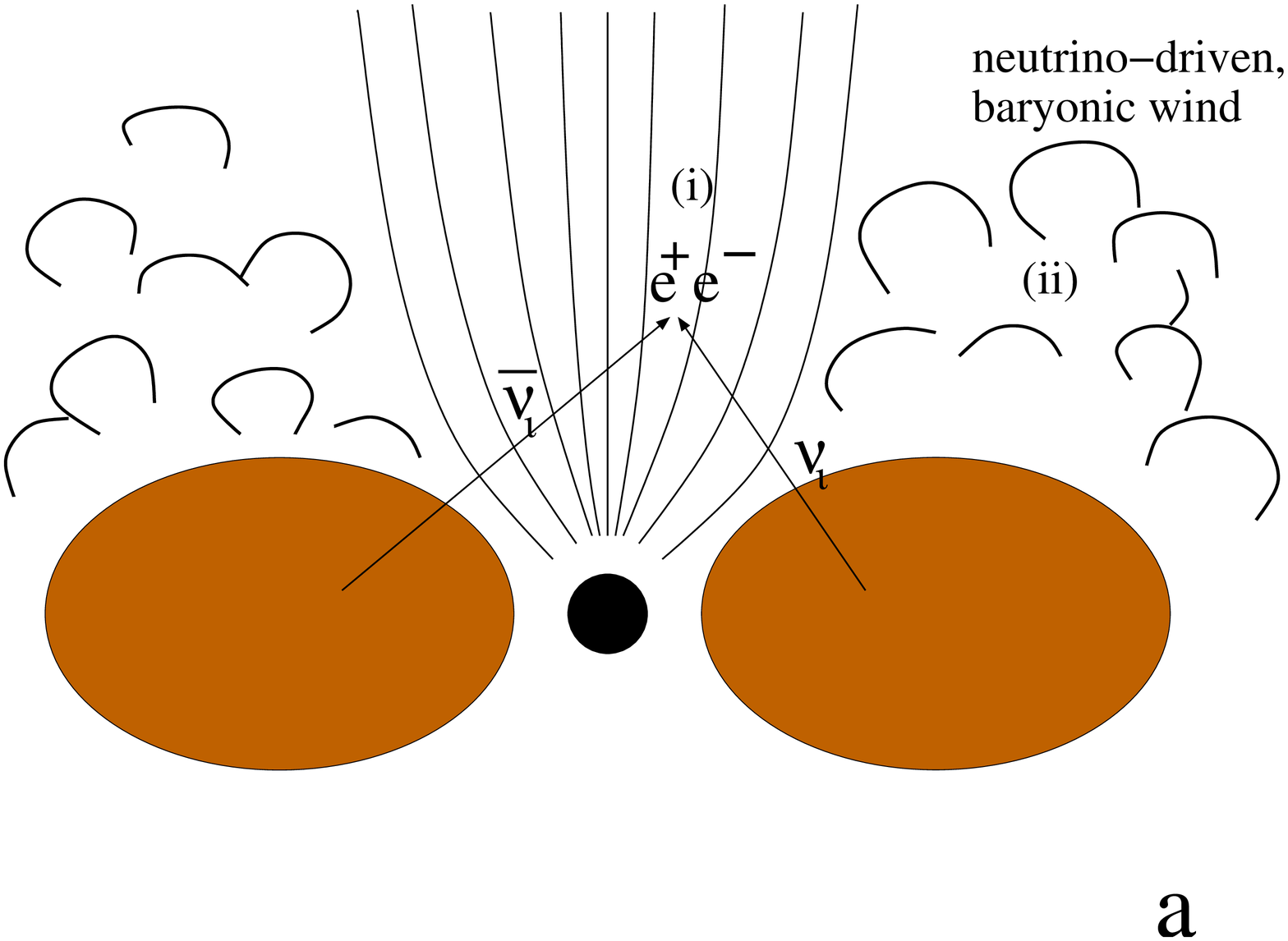,width=8cm,angle=0}\hspace*{1cm}
\psfig{file=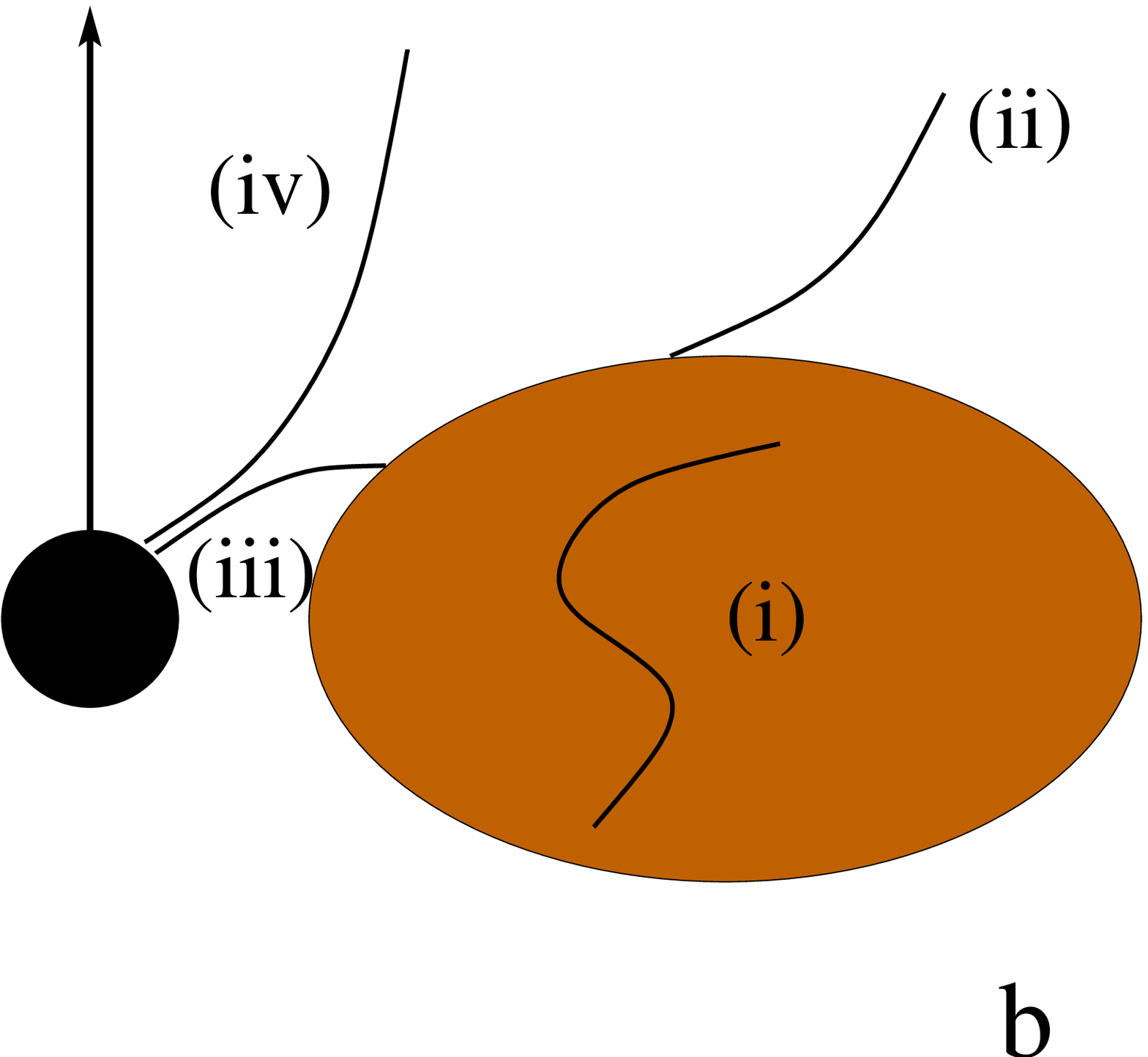,width=5cm,angle=0}}
\caption{\label{fig} Mechanisms to produce a GRB from a neutron star merger. 
{\bf Panel a}: neutrinos from the hot remnant annihilate and produce a relativistic
outflow (``fireball'', (i)). Moreover, the huge neutrino flux drives an
energetic baryonic wind (ii) that collimates the jet.
{\bf Panel b}: Strong magnetic fields anchored in the dense matter can convert 
the binding and/or spin energy into a Poynting outflow. Dynamo processes 
are believed to operate in accretion discs and physical
considerations suggest that fields generated in this way would have a
typical length-scale of the order of the disc thickness (i). Open field
lines that connect the disk to the outflow can drive a hydromagnetic wind 
(ii). The above mechanism can tap the binding energy of the
debris torus. At some stage after the merger a rapidly spinning black 
hole is expected to form. Its energy could be extracted in principle 
through MHD coupling to the rotation of the hole (iii and iv). 
Panel adapted from Blandford (2002).}
\end{figure}

\clearpage
\begin{figure*}
\centerline{\psfig{file=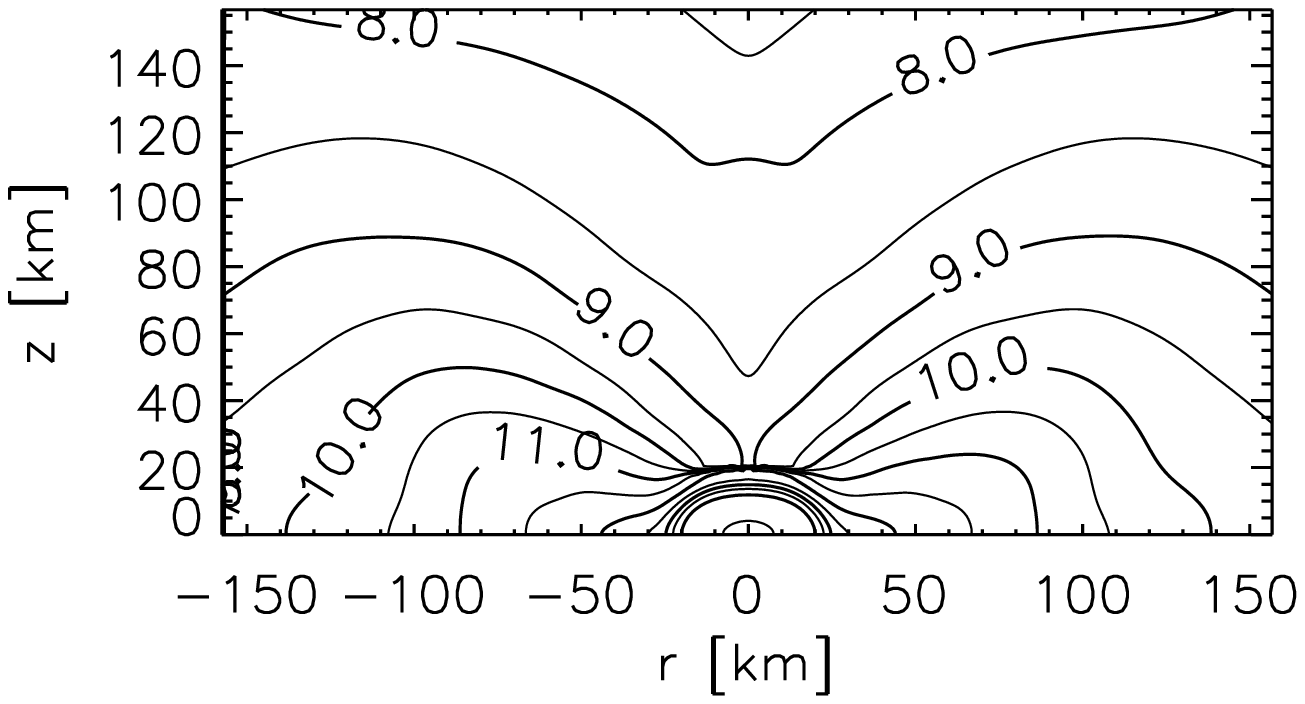,width=10cm,angle=0}
\psfig{file=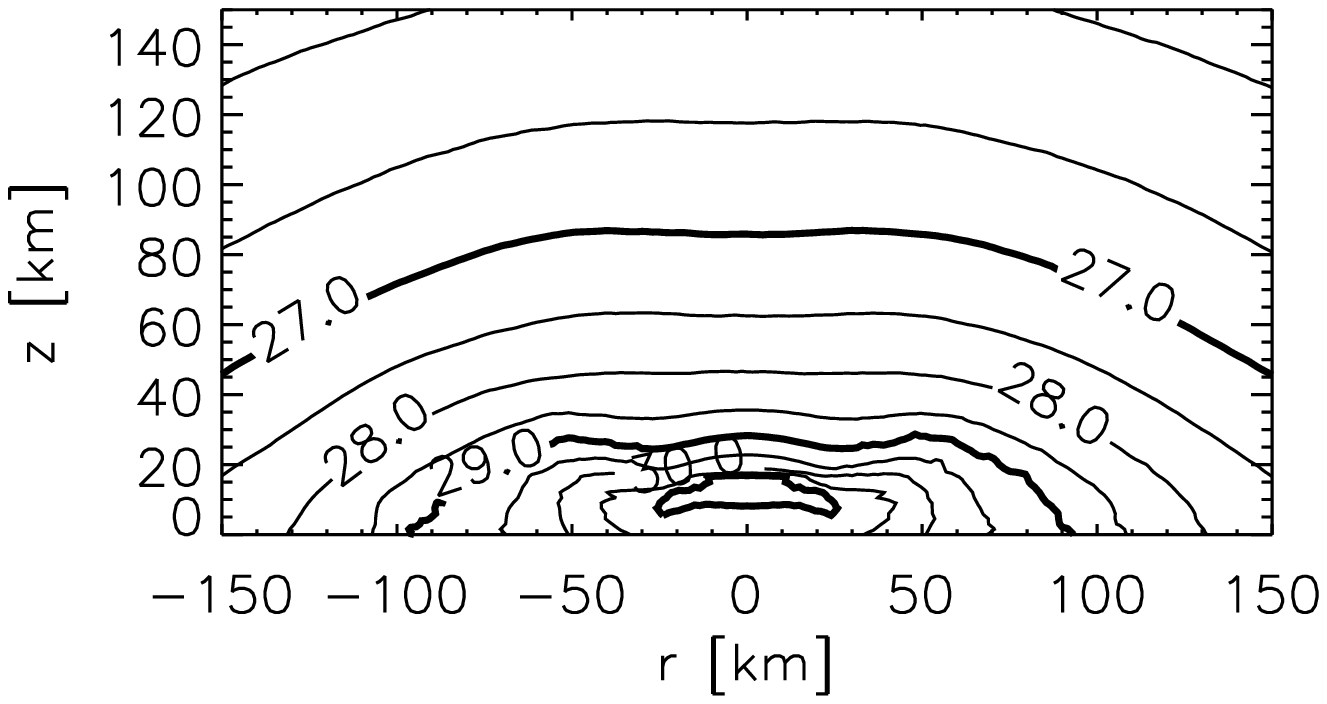,width=10cm,angle=0}}

\centerline{\psfig{file=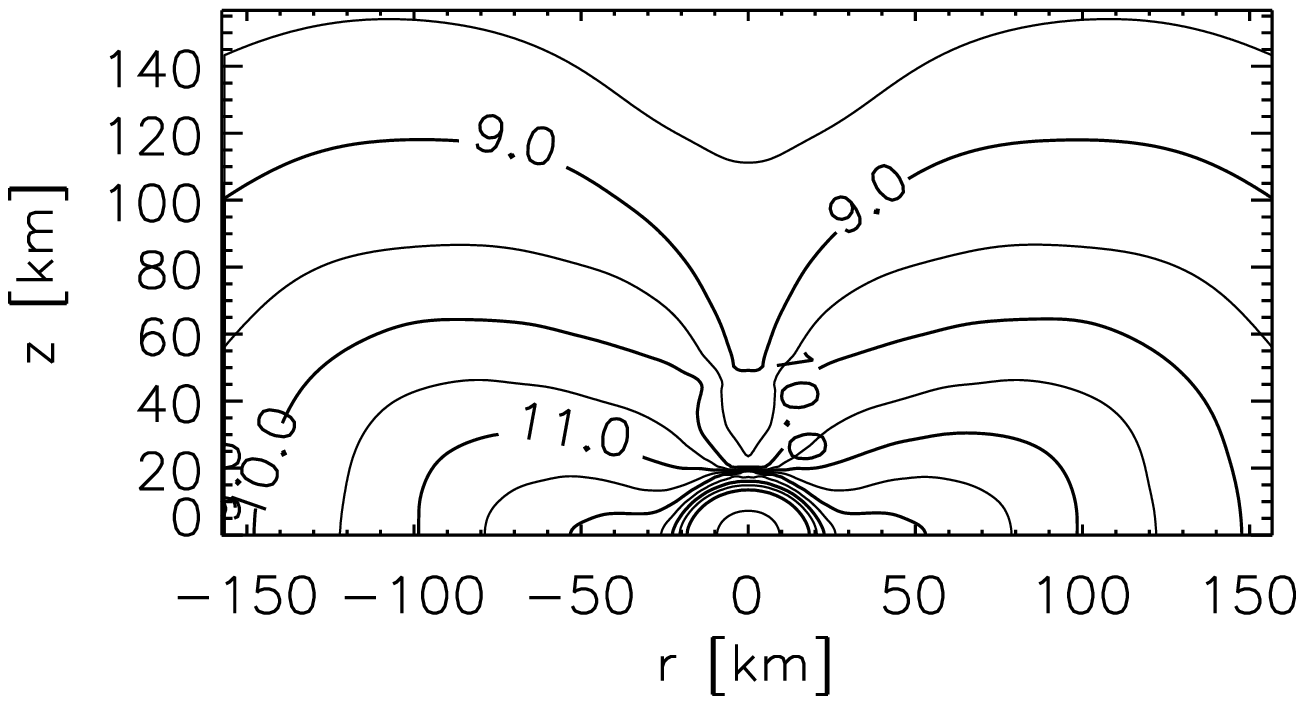,width=10cm,angle=0}
\psfig{file=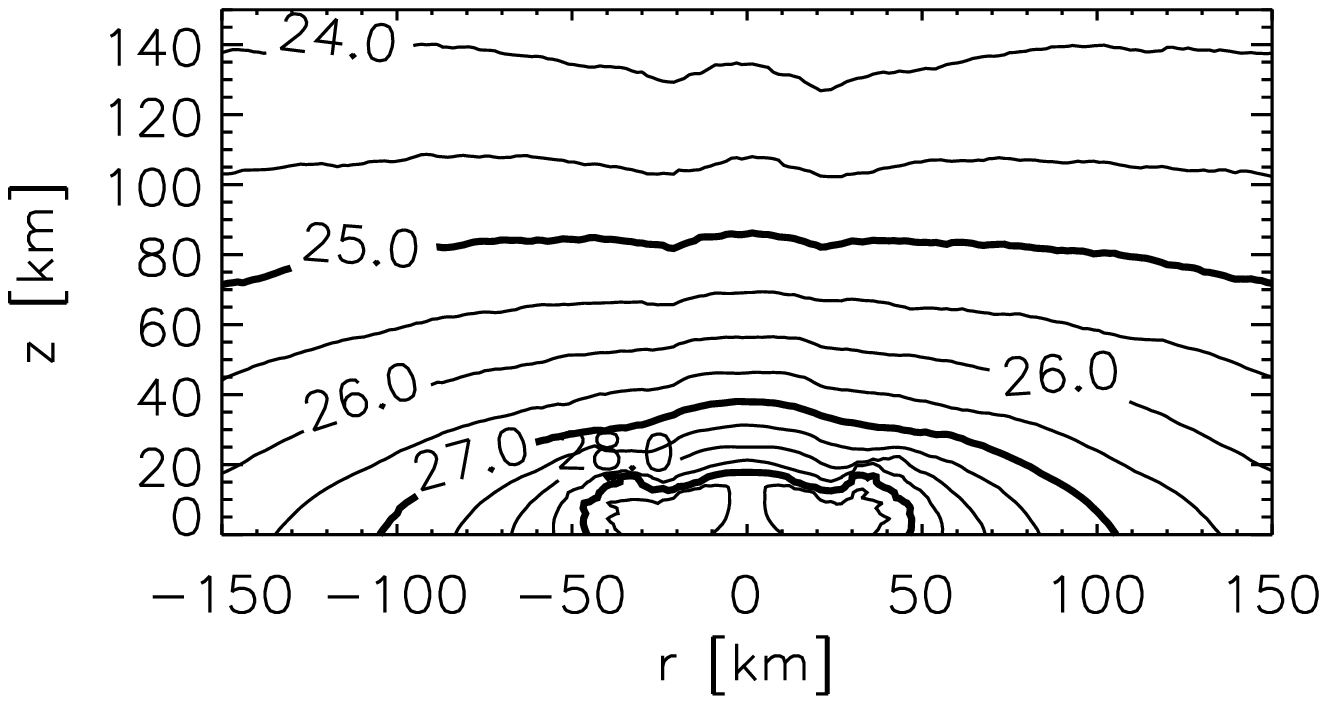,width=10cm,angle=0}}

\centerline{\psfig{file=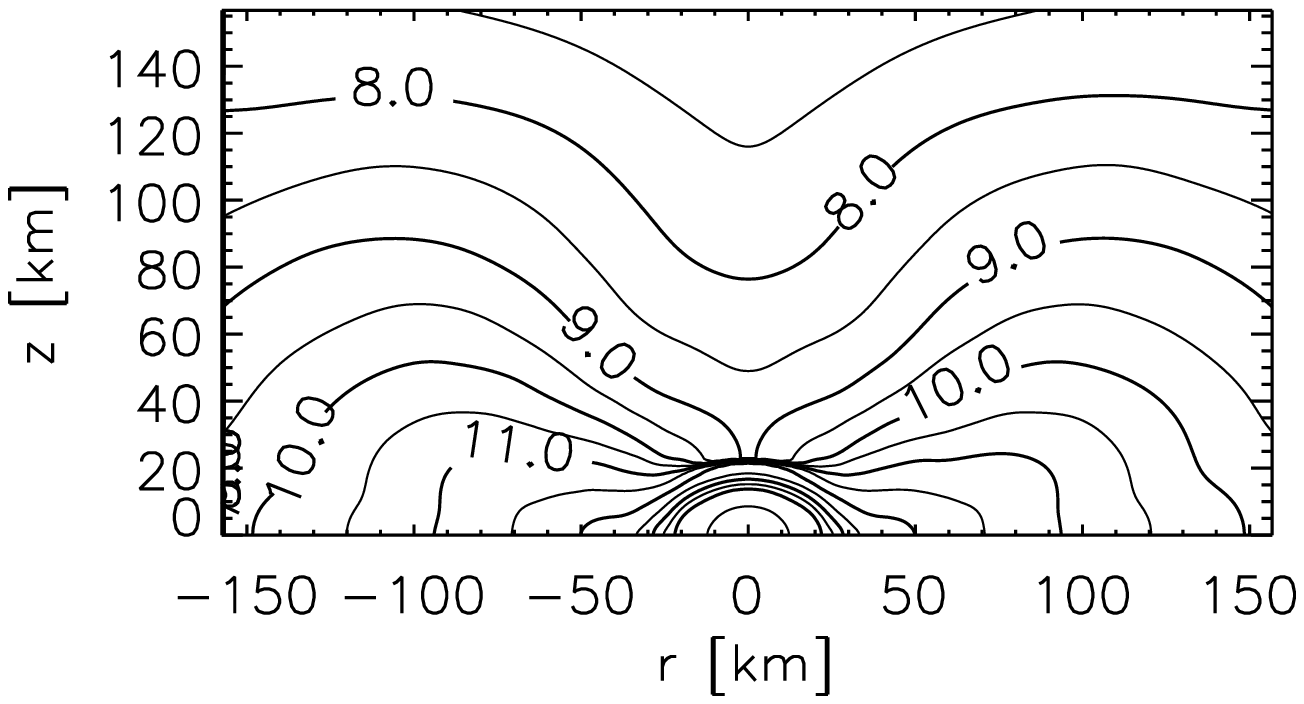,width=10cm,angle=0}
\psfig{file=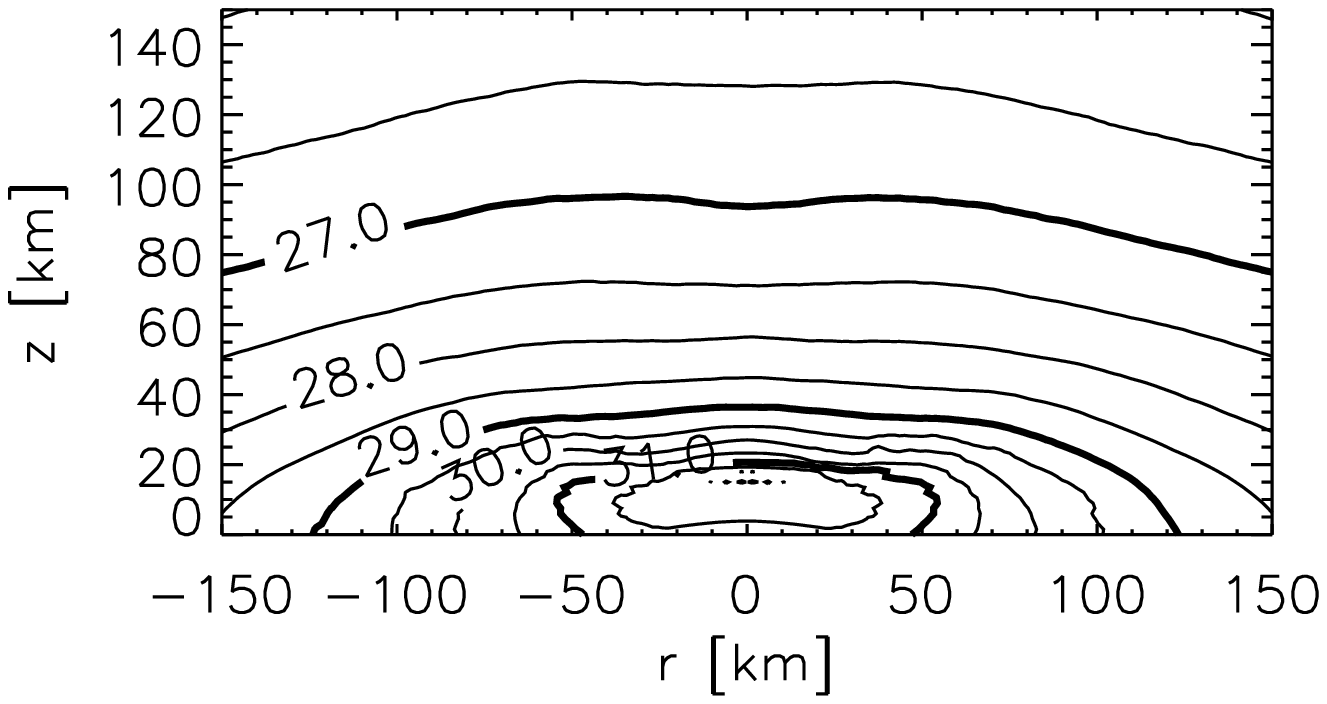,width=10cm,angle=0}}
\caption{\label{rho_rz} Left panel: densities along the rotation axis of 
run C, D and E (top to bottom). Right panel: contours of the energy deposited per time and volume (ergs/s cm$^3$) via $\nu\bar{\nu} \rightarrow e^{+}e^{-}$.}
\end{figure*}

\clearpage
\begin{figure}
\hspace*{1cm}\centerline{\psfig{file=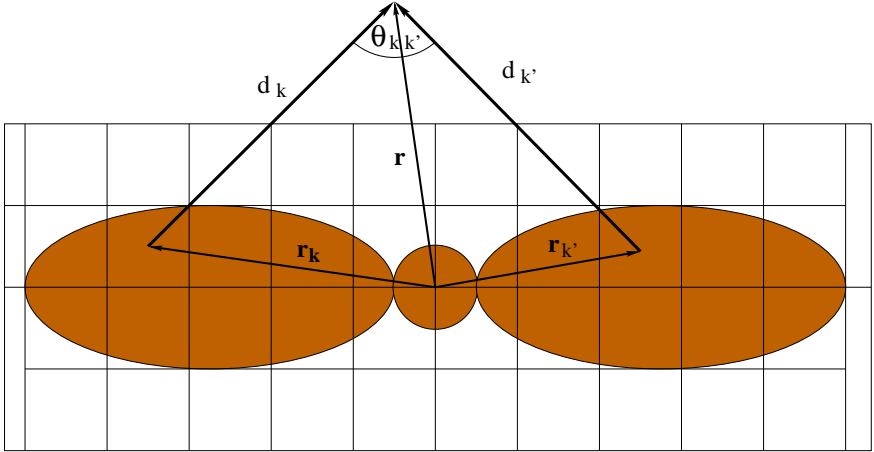,width=12cm,angle=-90}}
\caption{\label{vecs} Quantities from eq. (1): sketched is the grid with
the vectors needed to evaluate the annihilation at point $\vec{r}$ resulting
from neutrinos coming from cell $k$ and anti-neutrinos from cell $k'$.}
\end{figure}

\clearpage
\begin{figure}
\hspace*{-1cm}
\psfig{figure=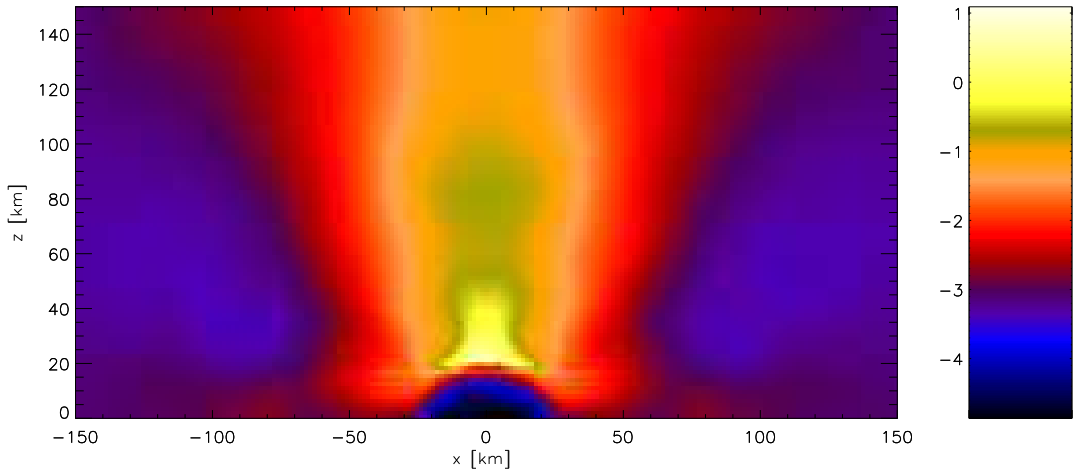,angle=90,width=15cm}
{\caption{Colour-coded is the ratio of energy deposited via 
$\nu\bar{\nu} \rightarrow e^{+}e^{-}$ to rest mass energy, $\eta$,
which is a measure of the maximum attainable Lorentz factor.
Shown are the values of log($\eta$) in the x-z-plane above the merged 
remnant of model C (no initial spins, 2 x 1.4 \msun) at the end of 
the simulation (t=18.3 ms). Due to the symmetry of the merger remnant
with respect to the orbital plane a similar jet will occur along the 
negative z-axis.}
\label{jetC}}
\end{figure}

\clearpage
\begin{figure}
\psfig{figure=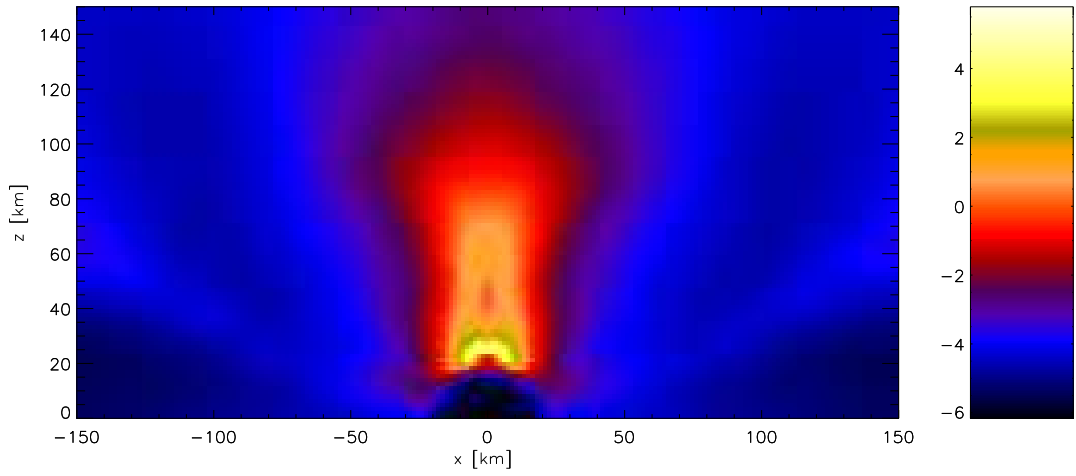,angle=90,width=15cm}
{\caption{Same as Figure \ref{jetC}, but for an initially coroting binary 
system, run F (see Table 1).}
\label{jetD}}
\end{figure}

\clearpage
\begin{figure}
\psfig{figure=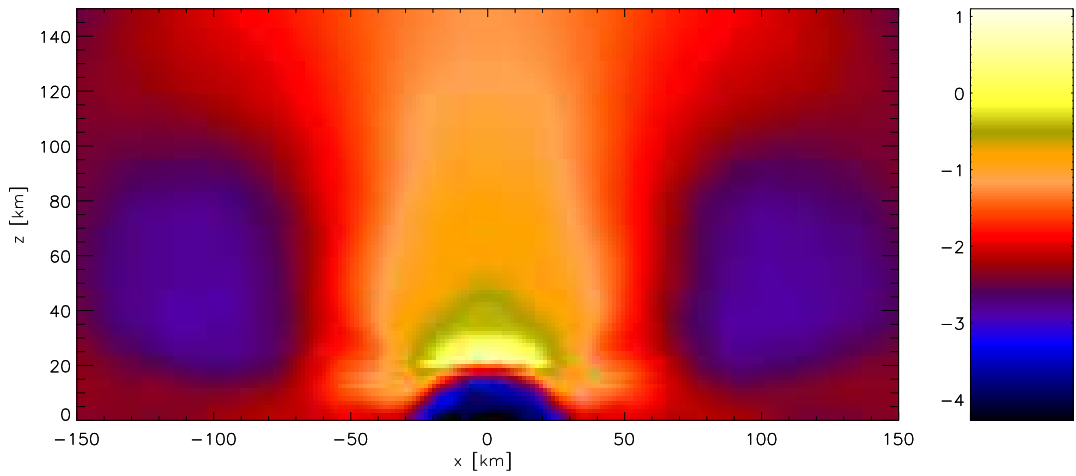,angle=90,width=15cm}
{\caption{Same as Figure \ref{jetC}, but for our extreme case 
(2x 2.0 \msun, no initial spins)}
\label{jetE}}
\end{figure}

\clearpage
\begin{figure}
\psfig{figure=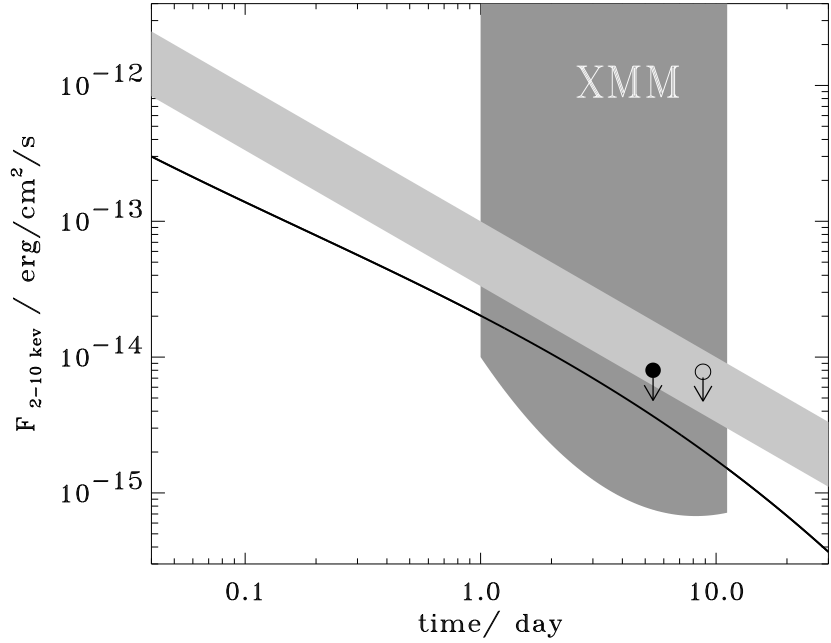,angle=0,width=15cm}
{\caption{Detectability of short GRBs produced via beamed $\nu\bar{\nu}$
annihilation (see text for details) with {\it XMM}. The
  solid line shows the 2-10 keV emission expected from afterglows of
  short GRBs. Model parameters have been chosen assuming that the
  relativistic shocks in short and long GRB afterglows have similar
  parameters except that the average kinetic energy per unit solid
  angle for short ones is $10^{51}/ 4\pi$ and the ejecta are
  collimated, with an initial aperture (half-angle) of 0.1 radians. A
  low density medium $n \sim 10^{-3} {\rm cm}^{-3}$ as expected in
  NS-NS merger models, has been chosen. The shaded region shows the
  extrapolated average afterglow emission at X-ray frequencies and
  later times, assuming a $\nu^{-1}$ spectrum (best fit to the
  measurement of Lazzati et al. 2001) and a $t^{-1}$ decay law.
 The black dot corresponds to GRB 020531, the open circle to GRB 011201.}
\label{xmm}}
\end{figure}

\clearpage
\begin{figure}
\psfig{figure=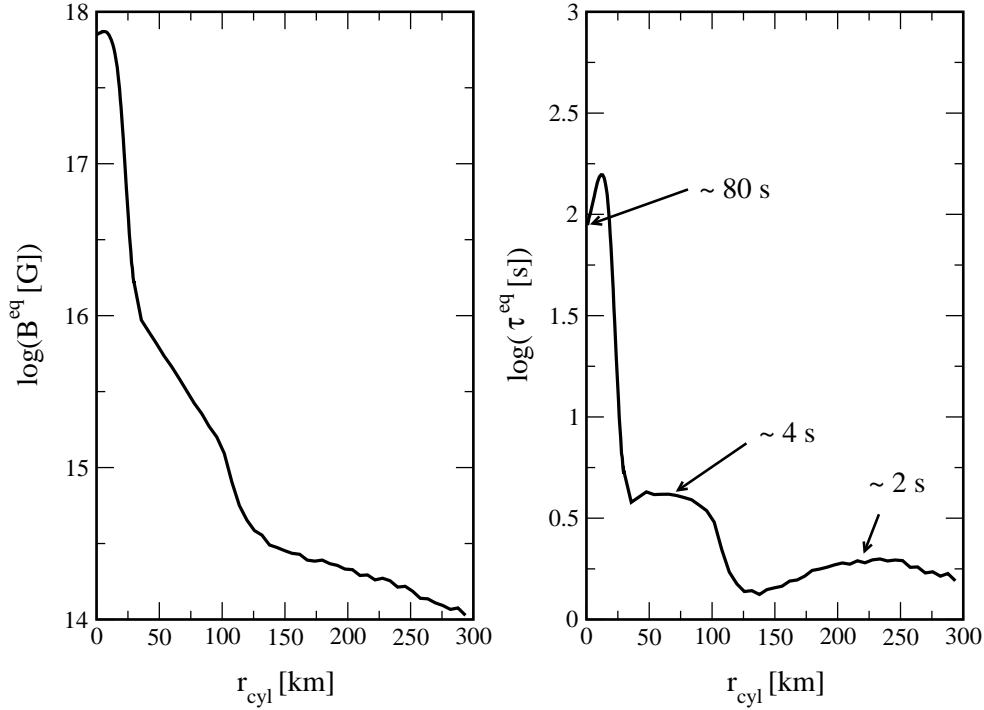,angle=-90,width=15cm}
{\caption{The left panel shows the equipartition magnetic field determined 
for our generic case, run C at 14.11 ms, from $B^{\rm eq}= \sqrt{8 \pi 
\rho c^2_s}$. The right panel shows the time it takes to reach the 
equipartition field strengths, starting with an initial field of 
$B_0= 10^{12}$ G. See text for details.}
\label{B_equipartition}}
\end{figure}

\clearpage
\begin{figure}
\psfig{figure=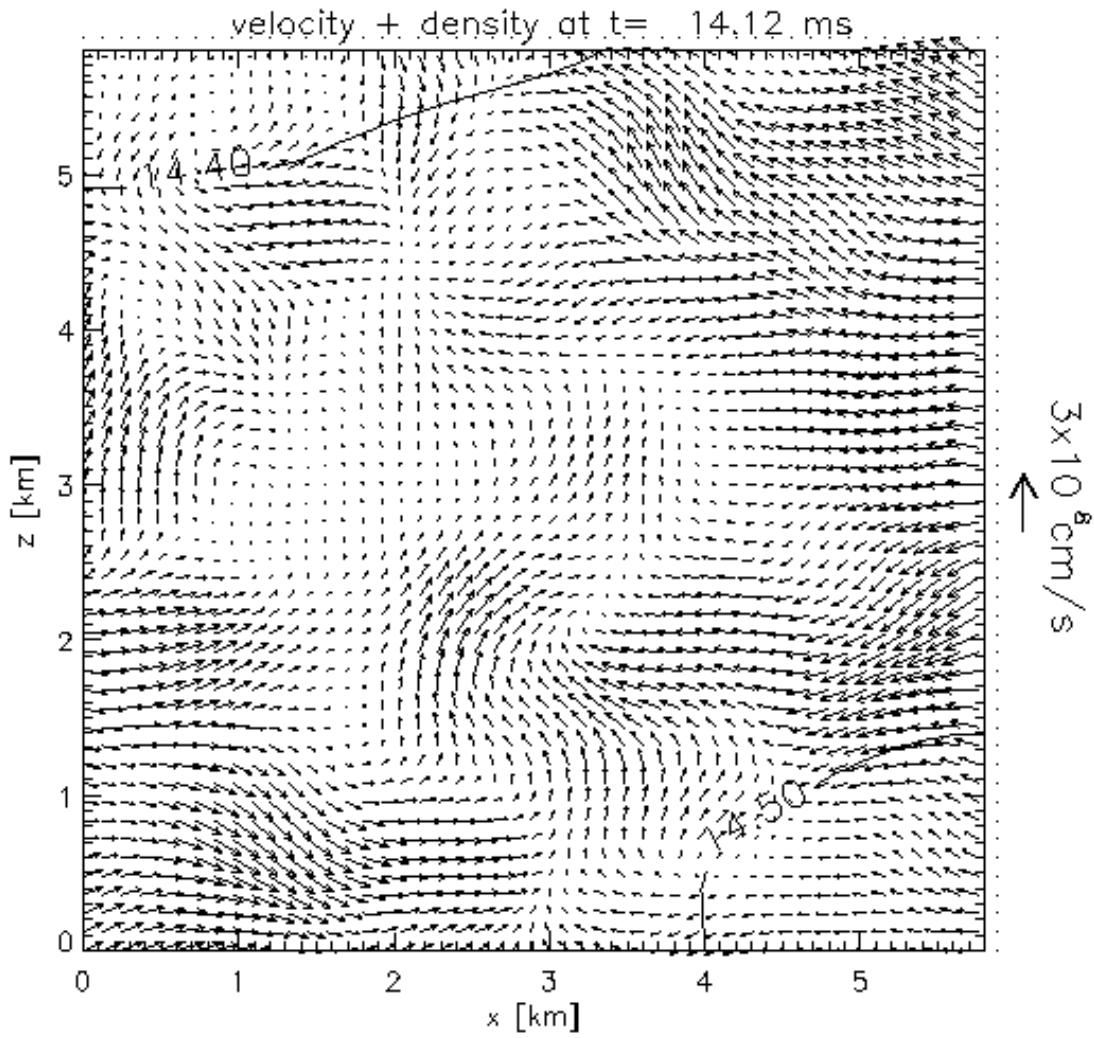,angle=0,width=15cm}
\caption{Velocity field inside the central, supermassive neutron star
of our generic case (2 x 1.4 \msun, no initial spin; run C).}\label{convection}
\end{figure}

\clearpage
\begin{figure}
\psfig{figure=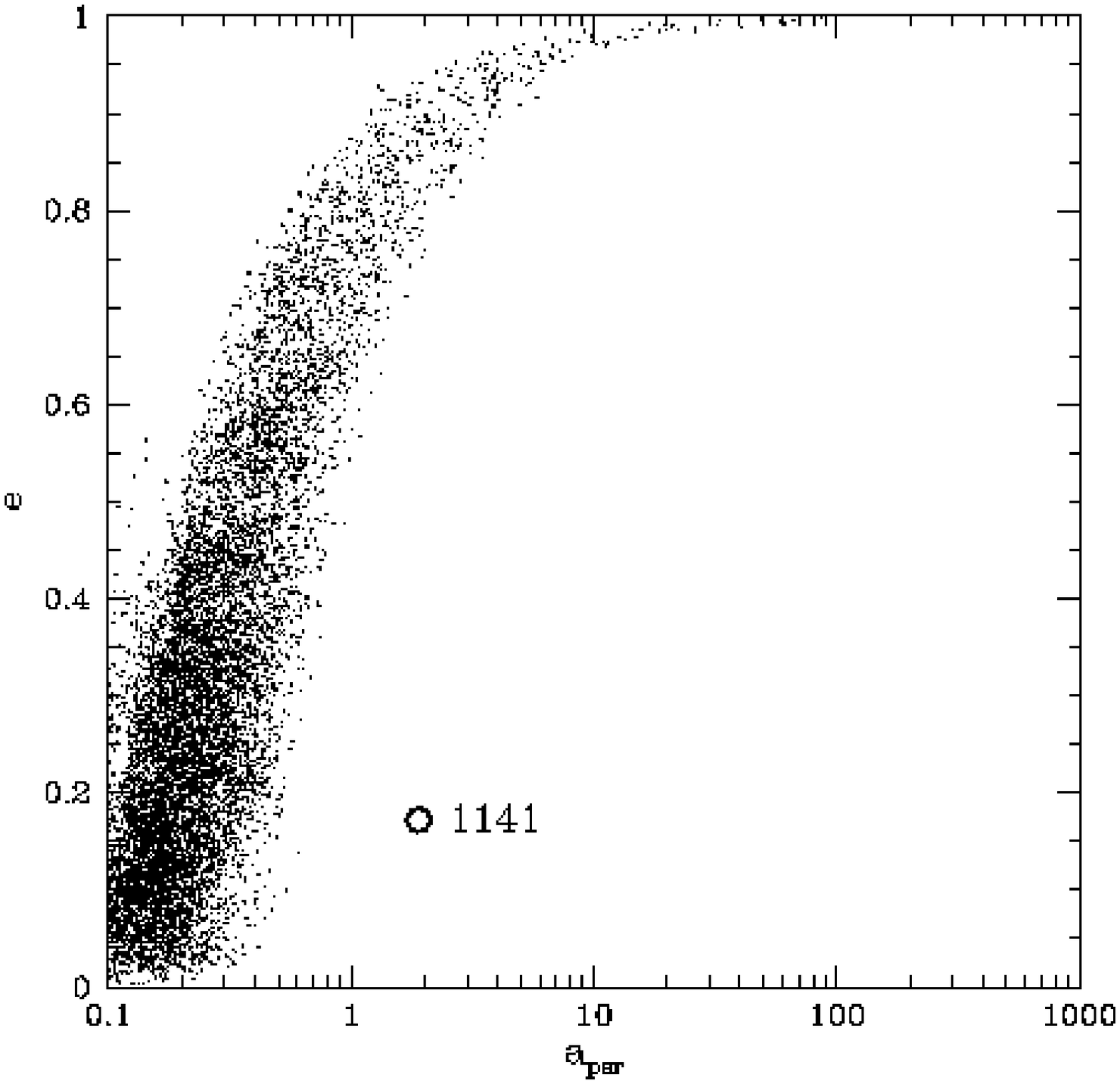,angle=0,width=15cm}
\caption{Plot of eccentricity $e$ as a function of separation (in solar
radii) for NS--NS systems undergoing a final common envelope phase
when the helium star fills its Roche lobe as described in the text.
In this plot, $\alpha_{\rm ce} \lambda_{\rm ce} = 1.0$ has been used.
The open circle is the observed (WD--NS) system
J1141--6545.}\label{mbd_fig1}
\end{figure}

\clearpage
\begin{figure}
\psfig{figure=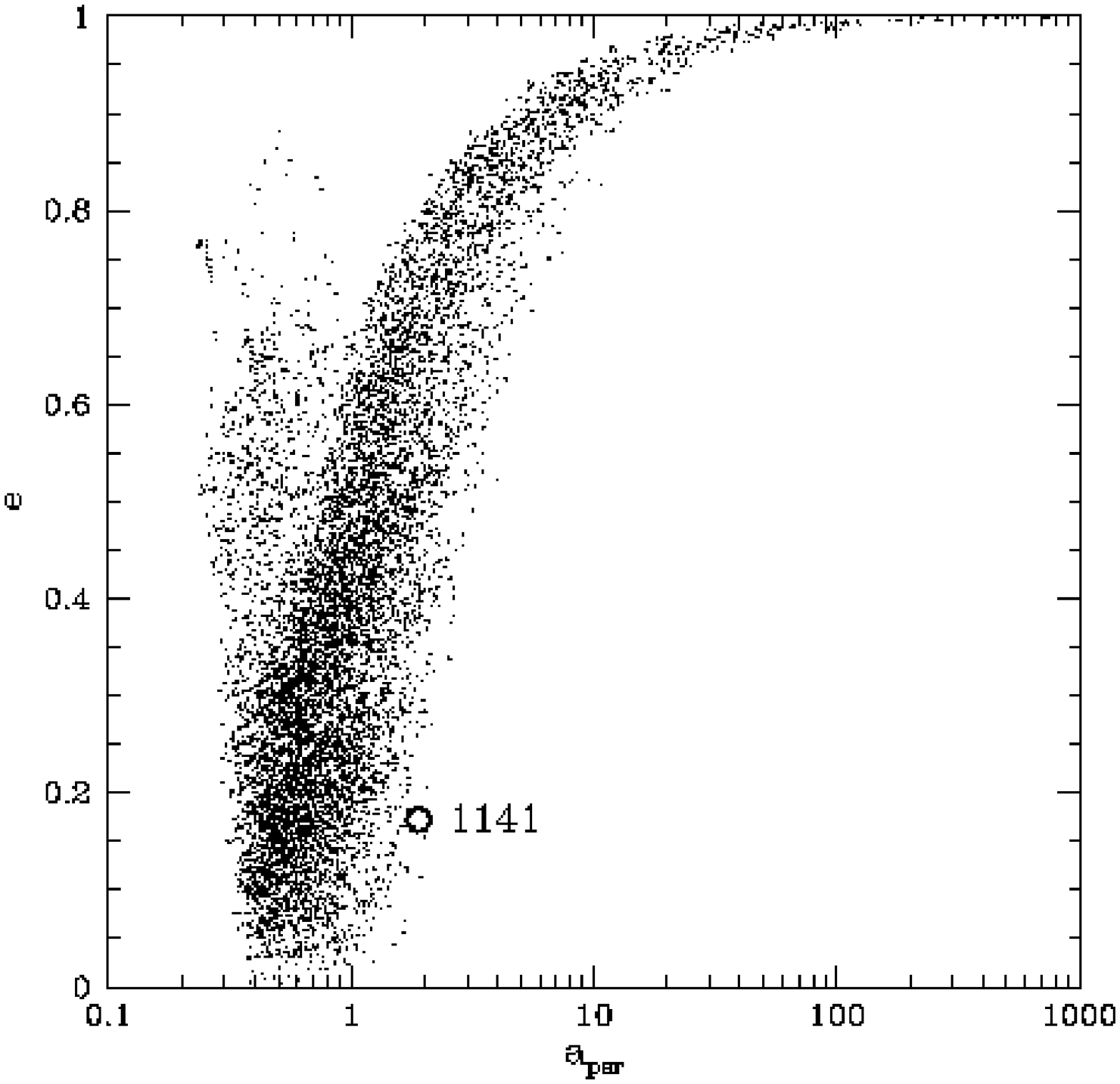,angle=0,width=15cm}
\caption{Plot of eccentricity $e$ as a function of separation (in solar
radii) for NS--NS systems undergoing a final common envelope phase
when the helium star fills its Roche lobe as described in the text.
In this plot, $\alpha_{\rm ce} \lambda_{\rm ce} = 10.0$.
The open circle is the observed (WD--NS) system
J1141--6545.}\label{mbd_fig2}
\end{figure}

\clearpage
\begin{figure}
\psfig{figure=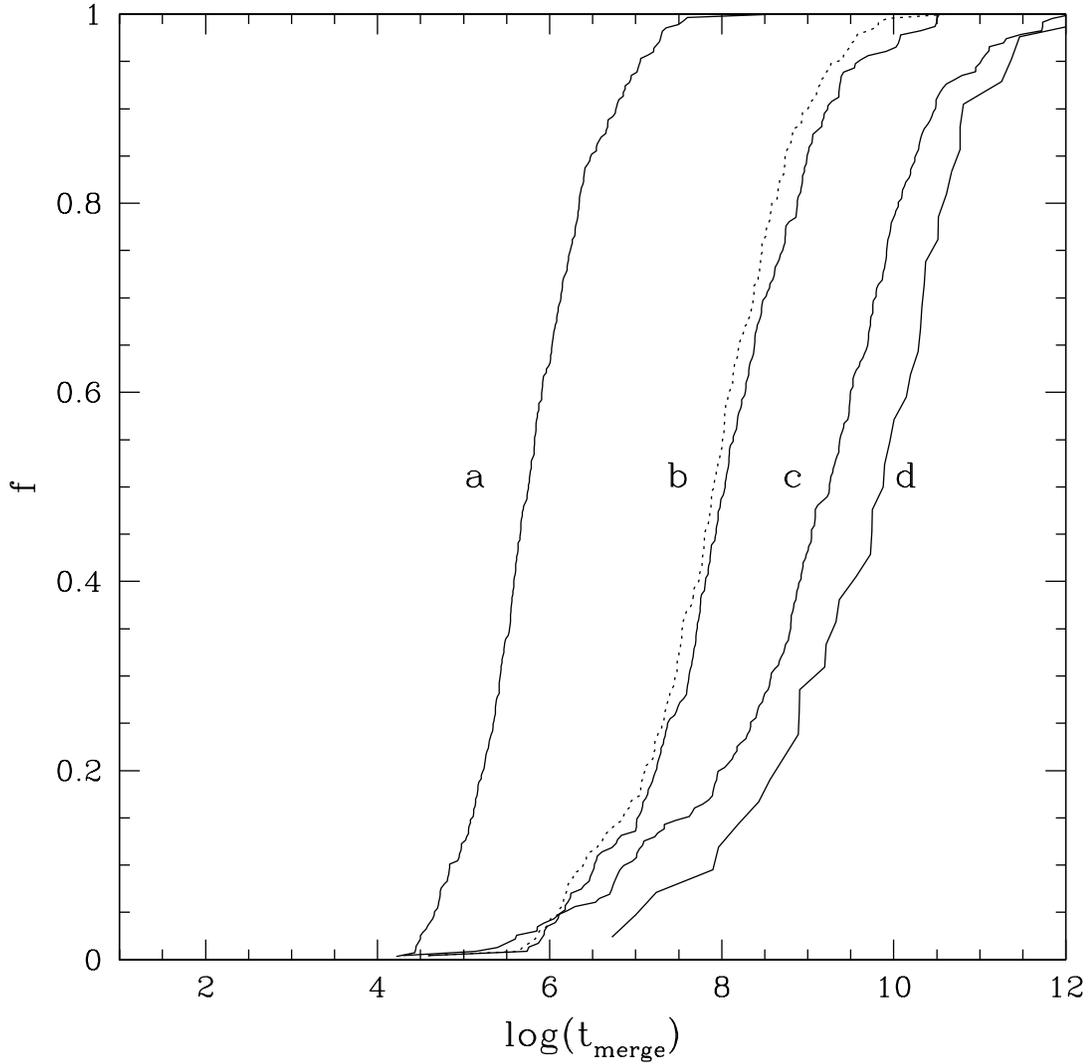,angle=0,width=15cm}
\caption{A plot of the cumulative distribution of merger times for
neutron star binaries under various assumptions for the formation route.
Considering first the four solid lines, from left to right:
common envelope phase ($\alpha_{\rm ce} \lambda_{\rm ce} = 1.0$) when 
helium star fills its Roche lobe (line a); 1141--like evolution (i.e.
Cyg X--2 mode of mass transfer occurs when helium star fills its Roche lobe;
line b); the distribution obtained
when the size of the helium star is neglected (line c); and the 
distribution obtained
for the systems following a 2303--like evolution (ie the helium star
fails to fill its Roche lobe and instead mass--loss occurs via a wind; line d).
The dotted line is produced assuming a common envelope phase but with
$\alpha_{\rm ce} \lambda_{\rm ce} = 10.0$.}\label{mbd_fig3}
\end{figure}

\clearpage
\begin{figure}
\psfig{figure=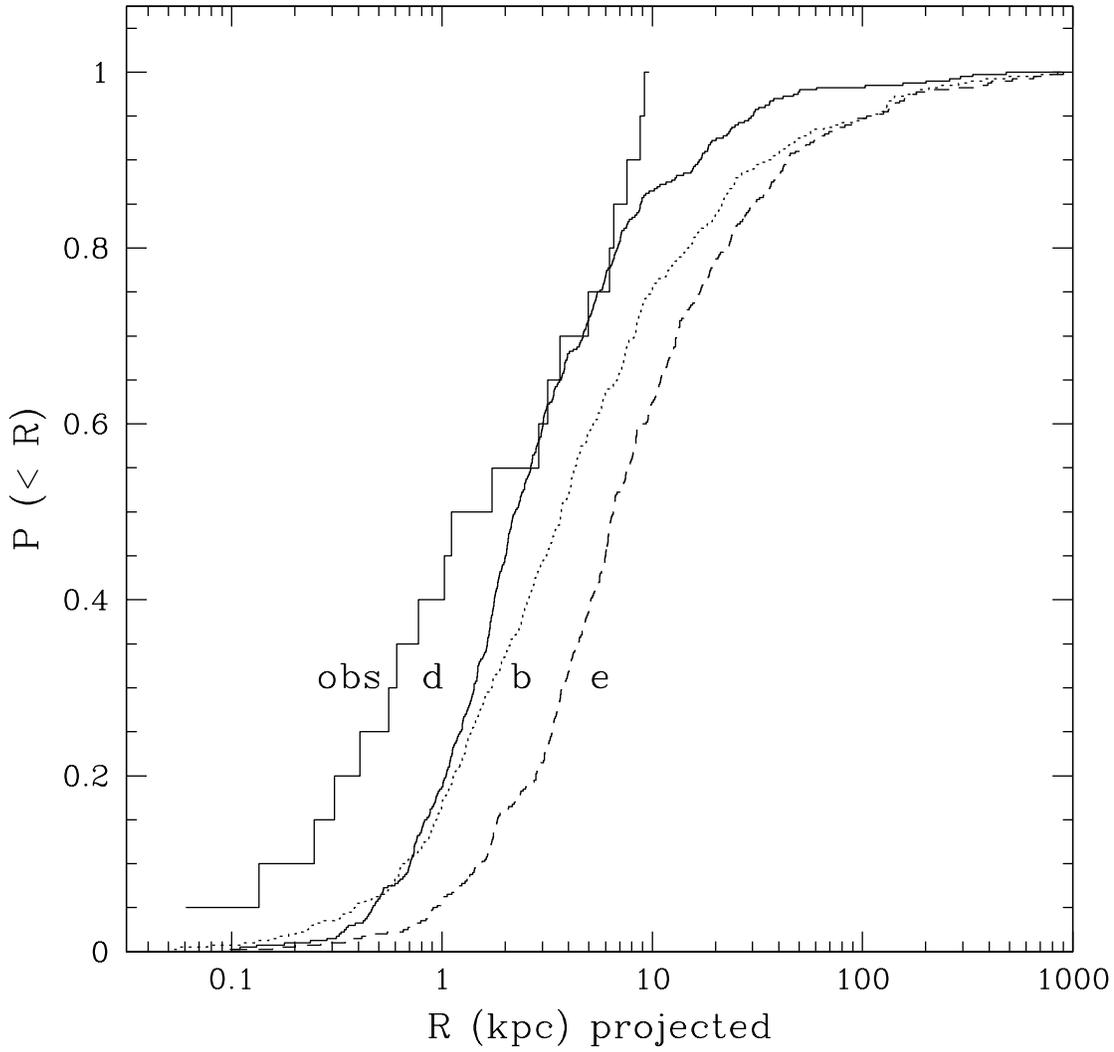,angle=0,width=15cm}
\caption{Distribution of the projected distances of the merger site 
from the galactic centre. This is shown 
for three galactic models: models d, b, and e (see Bloom et al. 1999 for
a discussion of the galactic models). 
We also plot the observed distribution from Bloom et al. 
(2002) (square, solid line).} \label{mbd_fig4}
\end{figure}

\end{document}